\documentclass{jpsj3}
\usepackage{amsthm,bm}

\usepackage{color}

\title{
Tilted-Cone Induced Cusps and Nonmonotonic Structures \\
in Dynamical Polarization Function of Massless Dirac Fermions
}
\author{Tatsuro \textsc{Nishine}$^{1}$
\thanks{E-mail address: nishine@slab.phys.nagoya-u.ac.jp}
, Akito \textsc{Kobayashi}$^{1,2}$
, and Yoshikazu \textsc{Suzumura}$^{1}$}

\inst{$^{1}$Department of Physics, Nagoya University, Furo-cho, Chikusa-ku, Nagoya 464-8602 \\
$^{2}$Institute for Advanced Research, Nagoya University, Furo-cho, Chikusa-ku, Nagoya 464-8602}

\abst{
 We examine  the dynamical property of the polarization function 
  for electrons with the tilted Dirac cone, considering the case where  
the  chemical potential measured from the contact point is finite, 
as found in the organic conductor  $\alpha$-(BEDT-TTF)$_2$I$_3$ 
(BEDT-TTF=bis(ethylene-dithio)tetrathiafulvalene) at ambient pressure.
Using the tilted Weyl equation with 
 the analytical treatment of the particle-hole excitation,
we show
 that the polarization function as a function of both the frequency and the momentum 
exhibits cusps and nonmonotonic structures. 
The polarization function depends not only on the magnitude 
but also on the direction of the external momentum. 
These properties are characteristic of the tilted Dirac cone, 
and are in contrast to those in the isotropic case of graphene. 
Furthermore, the results are applied to calculate 
the optical conductivity,  plasma frequency, and the screening of Coulomb interaction, 
which are also strongly influenced by the tilted cone. 
}
\kword{
Dirac cone, tilted Weyl equation, zerogap state, $\alpha$-(BEDT-TTF)$_2$I$_3$, 
polarization function, optical conductivity, plasma frequency
}

\begin{document}

\maketitle

\section{Introduction} 
The recent discovery of graphene\cite{graphene} has attracted much attention
 in the field of condensed matter because graphene exhibits massless Dirac fermions.
 Using the Weyl equation, \cite{McClure,Slonczewski} which describes the motion of massless Dirac fermions, several anomalous electronic properties have been investigated for a long time.\cite{Castro,Ando2005JPSJ} 
Another massless Dirac fermion is found in the quasi-two-dimensional  organic conductor $\alpha$-(BEDT-TTF)$_2$I$_3$
(BEDT-TTF=bis(ethylene-dithio)tetrathiafulvalene)
 under pressure 
\cite{Kajita1992f,Tajima2000f}. 
The existence was demonstrated theoretically
\cite{Kobayashi2004JPSJ73,Katayama2006JPSJ} using the band calculation, which is based on the transfer integrals estimated from the X-ray structure analysis.\cite{Kondo} Such an energy band  has  been confirmed by first-principles calculations.\cite{Ishibashi,Kino}
This novel state elucidates a long standing problem of anomalous phenomena observed in the conductor under pressure.\cite{KajitaFirst,Tajima2006JPSJ}

The massless Dirac fermion in the organic conductor is expected to exhibit a noticeable property due to the anisotropic velocity of the tilted Dirac cone, which is described by the tilted Weyl equation.\cite{Kobayashi2007JPSJ,Goerbig2008,Katayama2009,Kobayashi2009_rev}
It has been shown that  the tilt affects the characteristic temperature dependence of the Hall coefficient\cite{Kobayashi2008JPSJ}  
 where the magnitude of the tilt is characterized by the parameter 
 $\alpha =v_0 /v_c$ ($0 \le \alpha < 1 $), 
the ratio of the tilting velocity to the cone velocity. 
A recent experiment on  the transport phenomena of 
 $\alpha$-(BEDT-TTF)$_2$I$_3$ 
 suggested that $\alpha \cong 0.8$ using theoretical evaluation.\cite{TajimaRecen}
Since  the conductor is a layered two-dimensional massless Dirac fermion system,\cite{Osada2008,Tajima2009PRL}
 the interplane magnetoresistance also exhibits   noticeable 
 properties, as shown theoretically by the angle dependence of the  magnetic field.
\cite{Morinari2009JPSJ}  Moreover, new phenomena induced by the tilted Dirac cones have been maintained by calculating the transport coefficient under strong magnetic field, the electric-field-induced lifting of the valley degeneracy,\cite{Goerbig2009EPL} and the easy-plane pseudo-spin ferromagnet leading to 
 the  Kosterlitz-Thouless transition.\cite{Kobayashi2009}
In addition to the above studies, 
   dynamical properties such as electron-hole excitation and  collective excitation  are  promizing ingredients for verifying the role of tilting.
 Although the electronic state has been studied 
 extensively,  the dynamical properties associated with the polarization function 
 are  not yet clear enough  compared with those in the isotropic case of graphene.

In the present paper,
 we  analytically examine the dynamical polarization function with 
 the arbitrary wavevector and frequency, 
 and compare it with that in the  isotropic case of 
graphene.\cite{Kenneth,Ando2006,Wunsch2006,Hwang2007,Roldan}
We examine the metallic state  where 
  the contact point of the Dirac cone is located below the Fermi energy 
 as expected for the organic conductor $\alpha$-(BEDT-TTF)$_2$I$_3$.
\cite{Tajima2000f,Katayama2006JPSJ,Kondo} 
In \S 2, formulation for the polarization function is given.  In \S 3, the analytical expression  of the imaginary part is calculated, while  
 the real part is estimated semianalytically using the Kramers-Kronig relation. In terms of these results, we demonstrate  the cusps and nonmonotonic structures as characteristic of tilted Dirac cone.   
 On the basis of the results of \S 3, we calculate 
 the optical conductivity, plasma frequency, and screening of the Coulomb interaction in \S 4. The conclusion is given in \S 5.

\section{Formulation}
We consider the zero-gap state in $\alpha$-(BEDT-TTF)$_2$I$_3$, 
which has two tilted Dirac cones.
\cite{Katayama2006JPSJ} 
Among two contact points, $\pm \bm{k}_0$, corresponding to two valleys of cones, 
we focus on one of them, which is given by the state located close to 
$\bm{k}_0 (= k_{0x},k_{0y})$ with $k_{0x} < 0$ and $k_{0y} >0$. 
For such a state, the effective Hamiltonian is expressed as 
\cite{Kobayashi2007JPSJ}
\begin{align}
H=\sum_{\gamma \gamma'= 1,2}
\left(H_{\bm k}\right)_{\gamma \gamma'}
a_{\gamma,{\bm k}}^\dagger a_{\gamma',{\bm k}},
\label{eq:Hamiltonian}
\end{align}
which gives the tilted Weyl equation. 
The quantity $a_{{\bm k}\gamma}^\dagger$ denotes a creation operator of the electron 
where the momentum ${\bm k}$ is measured from that of the contact point. 
The apex of the Dirac cone for the conduction band touches  that of the valence band. 
Using the Luttinger-Kohn representation, 
where the two states at the contact point are chosen as the basis with the index $\gamma$, 
the matrix $H_{\bm k}$ is expressed as 
\begin{align}
H_{\bm k}=
{\bm k}\!\cdot\!
\begin{pmatrix}
{\bm v}_0 & {\bm v}_1-i{\bm v}_2 \\
{\bm v}_1+i{\bm v}_2 & {\bm v}_0
\end{pmatrix}.
\label{eq:Hmatrix}
\end{align}
Two states of the basis are chosen so as to give 
${\bm v}_0=(v_0,0),{\bm v}_1=(v,0),{\bm v}_2=(0,v')$ with $v =  v' = v_c$ 
(we use $\hbar = 1$). 
Diagonalizing eq.~(\ref{eq:Hmatrix}), the energy around the contact point is obtained as 
\begin{align}
\xi_{s{\bm k}}=v_0k_x+sv_c|{\bm k}|,
\quad
(s=\pm 1), 
\label{eq:disp}
\end{align}
where the first term comes from the tilting along the $k_x$-direction. 
Defining a tilting parameter $\alpha(< 1)$ as 
\begin{align}
\alpha=\frac{v_0}{v_c},
\end{align}
eq.~(\ref{eq:disp}) is rewritten as 
\begin{align}
\xi_{s{\bm k}}=(s+\alpha\cos\theta_{\bm k})v_c k,
\label{eq:disp2}
\end{align}
where ${\bm k}=(k_x, k_y)$, $k=|{\bm k}|$ and $\tan\theta_{\bm k}=k_y/k_x$. 
\begin{figure}[tb]
\begin{center}
\includegraphics{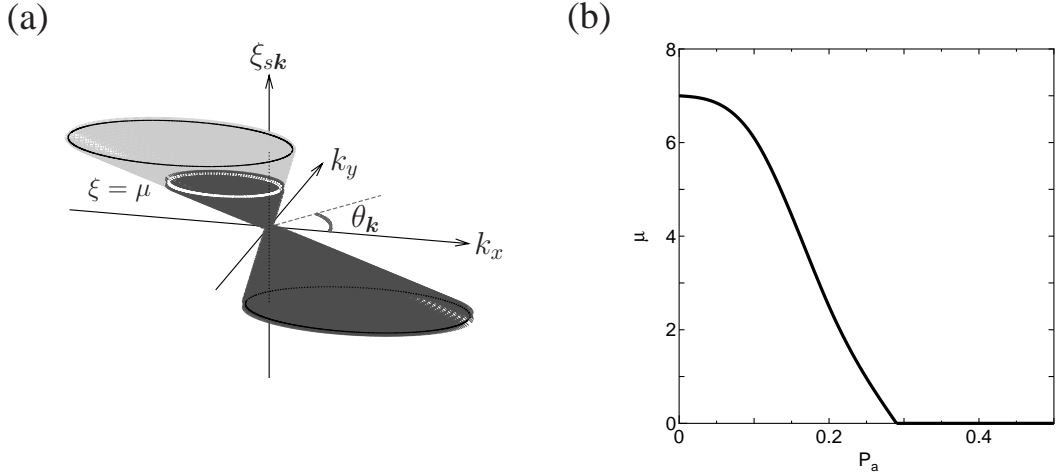}
\end{center}
\caption{
(a) Energy dispersion of the Dirac cone tilted along $k_x$-direction. 
The chemical potential is situated on the upper cone where the white circle denotes the Fermi surface. 
The angle $\theta_{\bm k}$, which is measured from the $k_x$-axis, 
is given by $\theta_{\bm k}=\arctan(k_y/k_x)$ in the $k_x$-$k_y$ plane.
(b)Chemical potential $\mu$ (meV) as a function of the uniaxial pressure $P_a$ (GPa) for $\alpha$-(BEDT-TTF)$_2$I$_3$.
}
\label{fig:cone}
\end{figure}
From eq.~(\ref{eq:disp2}), the trajectory for the fixed $\xi_{s,{\bm k}}$ 
gives the ellipse with a focus at $k=0$. 
Figure~\ref{fig:cone}(a) shows the energy dispersion obtained from eq.~(\ref{eq:disp}) 
where the Dirac cone of the upper band touches  that of the lower band at the contact point, $\bm{k}=0$. 
 In the present paper, the energy  is scaled by the Fermi energy
  (i.e., the chemical potential $\mu$), which  
 is measured from  the contact point. 
In Fig.~\ref{fig:cone}(b), 
\cite{Kondo}
 the  chemical potential of  $\alpha$-(BEDT-TTF)$_2$I$_3$ is shown
 as a function of the uniaxial pressure $P_a$ along the a-axis. 
 With increasing $P_a$ from ambient pressure, 
 $\mu$ decreases from 7 meV and reduces to zero at $P_a = 2.9$ kbar, above which the  zero gap state is obtained.   
The case of $\mu=0$ is briefly mentioned in the discussion.  
The eigenvector for eq.~(\ref{eq:disp}) is given by 
\begin{align}
{\bm F}_s(\bm k)=\frac{1}{\sqrt{2}}
\begin{pmatrix}
e^{-i\theta_{\bm k}} \\
s
\end{pmatrix}.
\end{align}
Thus, eq. (\ref{eq:Hamiltonian}) is rewritten as 
\begin{align}
\sum_{\gamma \gamma'= 1,2}
\left(H_{\bm k}\right)_{\gamma \gamma'}
a_{\gamma,{\bm k}}^\dagger  a_{\gamma',{\bm k}}
=
\sum_{s=\pm}\xi_{s,{\bm k}}w_{s,{\bm k}}^\dagger w_{s,{\bm k}},
\label{eq:Hamiltonian2}
\end{align}
where 
\begin{align}
\label{eq:transform}
w_{s,{\bm k}}
=
\sum_{{\bm k}}\left({\bm F}_s({\bm k})\right)^t_\gamma a_{\gamma,{\bm k}}.
\end{align}
For these two operators, Green's functions of the single particle are defined by 
\begin{subequations}
\begin{align}
\label{eq:GreenFunction}
G_s({\bm k}, {\rm i}\varepsilon_n)
=
-\frac{1}{2}\int_{-1/T}^{1/T}\hspace{-1.2em}{\rm d}\tau
\left< 
T_\tau w_{s,{\bm k}}(\tau)w_{s,{\bm k}}(0)^\dagger
\right> 
{\rm e}^{-{\rm i}\varepsilon_n \tau},
\\
G_{\gamma\gamma'}^{\rm LK}({\bm k},{\rm i}\varepsilon_n)
=
-\frac{1}{2}\int_{-1/T}^{1/T}\hspace{-1.2em}{\rm d}\tau
\left<
T_\tau a_{\gamma,{\bm k}}(\tau)a_{\gamma',{\bm k}}(0)^\dagger
\right>
{\rm e}^{-{\rm i}\varepsilon_n \tau},
\label{eq:GreenFunction_LK}
\end{align}
\end{subequations}
where $\varepsilon_n (={\rm i}(2n+1)\pi T)$ is the Matsubara frequency for 
the fermion. 
$T$ is the temperature and $T_\tau$ is the ordering operator for the imaginary time 
(we use $k_{\rm B}=1$) .
Equation~(\ref{eq:GreenFunction_LK}) is written explicitly as 
\begin{align}
G^{\rm LK}
=
\frac{1}{({\rm i}\varepsilon_n-v_0 k_x)^2-v_c^2 k^2}
\begin{pmatrix}
{\rm i}\varepsilon_n-v_0 k_x & v_c(k_x-{\rm i}k_y) \\
v_c(k_x+{\rm i}k_y) & {\rm i}\varepsilon_n-v_0 k_x
\end{pmatrix} .
\label{eq:GLK}
\end{align}
The polarization function per valley is calculated as 
\begin{align}
\Pi({\bm q},{\rm i}\omega_m)
=&
-2\sum_{\gamma \gamma'}\int_{0}^{1/T}\hspace{-0.5em}d\tau
{\rm e}^{i\omega_{m}\tau}
\big<T_{\tau}\rho_{\gamma,{\bm q}}(\tau)\rho^{\dagger}_{\gamma',{\bm q}}(0)\big>
\nonumber\\
=&
-\frac{2}{\beta L^2}\sum_{{\bm k}n \gamma\gamma'}
G_{\gamma\gamma'}^{\rm LK}({\bm k}+{\bm q},{\rm i}\varepsilon_n+{\rm i}\omega_m)
G_{\gamma'\gamma}^{\rm LK}({\bm k},{\rm i}\varepsilon_n)
\nonumber\\
=&
-\frac{2}{\beta L^2}\sum_{{\bm k}nss'}
\big|{\bm F}^{\dagger}_{s}({\bm k}){\bm F}_{s'}({\bm k}+{\bm q})\big|^2
G_{s'}({\bm k}+{\bm q},{\rm i}\varepsilon_n+{\rm i}\omega_m)
G_{s}({\bm k},{\rm i}\varepsilon_n) , 
\label{eq:Pi_def}
\end{align}
where the freedom of the spin is included. 
${\rm i}\omega_m (={\rm i}2m\pi T)$ is the Matsubara frequency for the boson. 
$\rho_{\gamma,{\bm q}}$ is the electron density operator defined as
$\rho_{\gamma,{\bm q}}=\sum_{{\bm k}}a^{\dagger}_{\gamma,{\bm k}}a_{\gamma,{\bm k}+{\bm q}}$ and 
\begin{align}
\big|{\bm F}_s^{\dagger}(\bm k){\bm F}_{s'}({\bm k}')\big|^2
=
\frac{1}{2}\Bigl[1+ss'\cos(\theta_{\bm k}-\theta_{{\bm k}'})\Bigr] .
\label{eq:FF2}
\end{align}
After performing an analytical continuation given by $i \omega_m \longrightarrow \omega+i\eta, (\eta = + 0)$, one obtains  
\begin{align}
\label{eq:Polization}
\Pi({\bm q},\omega)
=&
\sum_{ss'}
\left(\frac{-2}{L^2}\right) \sum_{{\bm k}}
\big|{\bm F}^{\dagger}_{s}({\bm k}){\bm F}_{s'}({\bm k}+{\bm q})\big|^2
\frac{f(\xi_{s,{\bm k}})-f(\xi_{s',{\bm k}+{\bm q}})}
{\omega+{\rm i}\eta-(\xi_{s',{\bm k}+{\bm q}}-\xi_{s,{\bm k}})}
\nonumber\\
\equiv&
\sum_{ss'}\Pi_{ss'}({\bm q},\omega) ,
\end{align}
where $f(\xi) = (1 + {\rm e}^{(\xi - \mu)/T)})^{-1}$.
The term $\Pi_{ss}$ comes from the electron-hole excitation of the intraband, 
while the term $\Pi_{s, -s}$ comes from that of the interband.

In the present paper, we examine the case for $T=0$ 
and calculate eq.~(\ref{eq:Polization}) analytically by performing the integration over $\bm k$. 
The case for $ \omega \ge  0$ and  $ \mu \ge 0 $ is calculated, 
while the case for $\omega<0$ is obtained from 
\begin{align}
\Pi({\bm q},-\omega)=\bigl[ \Pi(-{\bm q},\omega) \bigr]^{*},
\label{eq:Pi_conj}
\end{align}
which comes from the property of $\rho_{\gamma,\bm{q}}$ of the tilted cone, i.e., 
without inversion symmetry. 
In the isotropic case with inversion symmetry,  
one obtains $\Pi({\bm q},-\omega)=\bigl[ \Pi({\bm q},\omega) \bigr]^{*}$.

Here, we note that, by taking account of the freedom of both spin and valley, 
the polarization function of the total system is given by 
\begin{align}
\Pi^{\rm total}({\bm q},\omega)= 
\Pi({\bm q},\omega) + \Pi(-{\bm q},\omega).
\label{eq:Pi_total}
\end{align}
The quantity of $\Pi({\bm q},\omega)$ per spin and valley 
is calculated in \S 3, while  
 charge response is calculated using $\Pi^{\rm total}({\bm q},\omega)$ in \S 4. 
\section{Polarization Function}

\subsection{Analytical calculation}
\begin{figure}[tb]
\begin{center}
\includegraphics{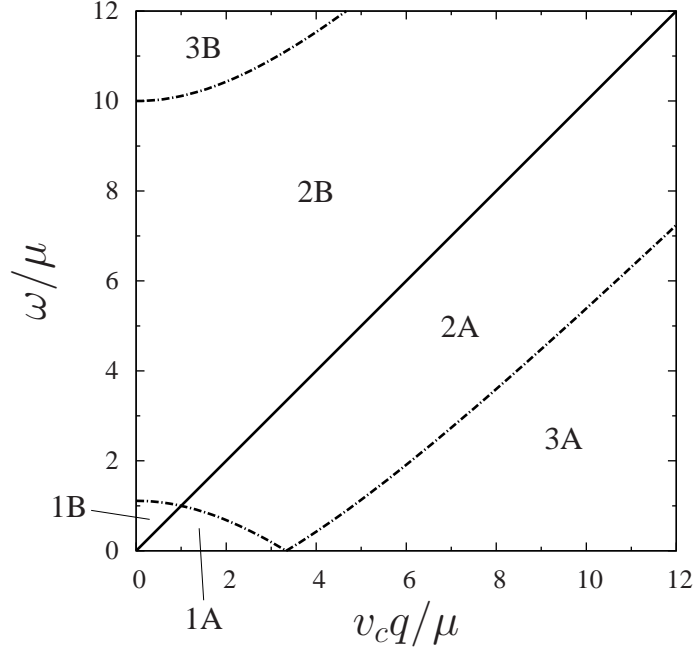}
\end{center}
\caption{
Several regions on the $q$-$\omega$ plane for ${\rm Im} \Pi(\bm{q},\omega)$ with the fixed $\theta_q=\pi/2$ and $\alpha=0.8$.
The solid line denotes  the resonance frequency, $\omega = \omega_{\rm res}$.
 The  dash-dotted line between 2B and 3B corresponds to  $\omega_{\rm B}$
 and the line between 1A and 2A (and also 1B and 2B)
  corresponds to $\omega_{\rm A}$.
 The imaginary part is absent in the regions of both 3A and 1B.
The boundary between 2A and 3A is given by  $\omega_+$.
The intersection point of $\omega_A$ and $\omega_{\rm res}$ is given by 
$v_cq/\mu = 1/(1+\alpha \cos \theta_{\bm{q}})$ and $\omega/\mu =1$
 while $\omega_A/\mu =2/(1+\alpha)$ at  $v_cq/\mu = 0$, and $\omega_A=0$ 
 at  $v_cq/\mu = 2/(\sqrt{(1-\alpha^2)(1-\alpha^2 \cos^2\theta_{\bm{q}})}$.
All the boundaries except for the solid line are followed by the cups, i.e, 
the jump of the derivative with respect to $q$ and $\omega$. 
  }
\label{ImP_region_2pi4}
\end{figure}

We  examine the imaginary part using eq.~(\ref{eq:Polization}).
By noting  ${\rm Im}\Pi_{--} = {\rm Im}\Pi_{+-} = 0$ 
 due to  the valence band  fully occupied, 
 we calculate the remaining parts of  ${\rm Im}\,\Pi_{++}$
 and ${\rm Im}\,\Pi_{-+}$, which correspond to the process  excited from the conduction band and valence band, respectively. 
After  performing the tedious but straightforward calculation, 
we obtain the analytical results (Appendices A and B): 
\begin{subequations}
\begin{align}
&
{\rm Im}\,\Pi_{++}(q,\theta_{\bm q},\omega)
=\sum_{\zeta = {\rm 1A,2A,3A}} \Pi''_{\zeta},
\label{eq:Pi_++}
\\
&
{\rm Im}\,\Pi_{-+} (q,\theta_{\bm q},\omega)
=\sum_{\zeta = {\rm 1B,2B,3B}} \Pi''_{\zeta}.
\label{eq:Pi_-+}
\end{align}
\end{subequations}
The results consist of  
six regions on the $q$-$\omega$ plane where the typical case with $\theta_q= \pi/2$ and $\alpha=0.8$ is shown in Fig. \ref{ImP_region_2pi4}.
These regions of 1A,  2A,  3A,  1B,  2B, and 3B 
  are classified into two regions, A and B, corresponding to the process of intraband and interband excitations, respectively.  The regions A and B are separated by a solid line expressed as  
\begin{align}
\omega_{\rm res} = (1 + \alpha \cos \theta_{\bm q})v_c q  , 
\label{eq:omega_res}
\end{align}
which is called the resonance frequency. 
 The resonance frequency is obtained owing to the nesting of the excitations with the linear dispersion, and the polarization function diverges with the chirality factor eq.(\ref{eq:FF2}) taking a maximum.\cite{Polini}
The boundary between 2A and 3A is given by $\omega_+$. 
The boundary between 2A and 1A ( 2B and 1B)  is given by 
 $\omega_{\rm A}$.
The boundary between 2B and 3B is given by 
 $\omega_{\rm B}$. These frequencies are calculated as  
\begin{align}
&
\frac{
\omega_{+}(\theta_{\bm q})
}{\mu}
=\alpha\frac{v_c q}{\mu}\cos\theta_{\bm q}-\frac{2}{1-\alpha^2}
+\sqrt{\left(\frac{v_c q}{\mu}\right)^{\!\!2}-\frac{4\alpha\frac{v_c q}{\mu}\cos\theta_{\bm q}}{1-\alpha^2}
+\left(\frac{2\alpha}{1-\alpha^2}\right)^{\!\!2}} ,
\label{eq:omega_+}
\\
&
\frac{
\omega_{\rm A(B)}(\theta_{\bm q})
}{\mu}
=\alpha\frac{v_c q}{\mu}\cos\theta_{\bm q}+\frac{2}{1-\alpha^2}
-(+)\sqrt{\left(\frac{v_c q}{\mu}\right)^{\!\!2}+\frac{4\alpha\frac{v_c q}{\mu}\cos\theta_{\bm q}}{1-\alpha^2}
+\left(\frac{2\alpha}{1-\alpha^2}\right)^{\!\!2}} .
\label{eq:omega_AB}
\end{align}
In the case of the isotropic Dirac cone (i.e., $\alpha = 0$),\cite{Wunsch2006,Hwang2007} there is a boundary given by $\omega/\mu + v_cq/\mu = 2$ which separates 1A and 2A (1B and 2B) for the intraband (the interband). In the regions 3A and 1B, the imaginary part vanishes. 
In the present case of the tilted Dirac cone, the boundary between 1A and 2A exhibits a noticeable behavior characterized by the appearance of cusps for the imaginary part
as shown in this section.


The real part is calculated from the imaginary part 
using the Kramers-Kronig relation:
\begin{align}
{\rm Re}\,\Pi(q,\theta_{\bm q},\omega)
=&
\frac{1}{\pi}{\cal P}\hspace{-0.5em}\int_{-\infty}^{\infty}\hspace{-1em}dx\,\,
\frac{{\rm Im}\,\Pi(q,\theta_{\bm q},x)}{x-\omega}
\nonumber\\
=&
\frac{1}{\pi}{\cal P}\hspace{-0.5em}\int_{0}^{\infty}\hspace{-1em}dx\,\,
\frac{{\rm Im}\,\Pi(q,\theta_{\bm q},x)}{x-\omega}
+\frac{1}{\pi}{\cal P}\hspace{-0.5em}\int_{0}^{\infty}\hspace{-1em}dx\,\,
\frac{{\rm Im}\,\Pi(q,\pi+\theta_{\bm q},x)}{x+\omega} \; ,
\label{eq:KK}
\end{align}
where we used, from  eq.~(\ref{eq:Pi_conj}), the relation 
 \begin{align}
&{\rm Re\,}\Pi({\bm q},-\omega)={\rm Re\,}\Pi(-{\bm q},\omega)
 ={\rm Re\,}\Pi(q,\pi+\theta_{\bm q},\omega) , 
 \nonumber \\
&{\rm Im\,}\Pi({\bm q},-\omega)=-{\rm Im\,}\Pi(-{\bm q},\omega)
=-{\rm Im\,}\Pi(q,\pi+\theta_{\bm q},\omega) .
\label{eq:sym}
\end{align}
The result of the semianalytical calculation 
 is given in  Appendix C

\subsection{Behavior of the polarization function}
\begin{figure}[tb]
\begin{center}
\includegraphics{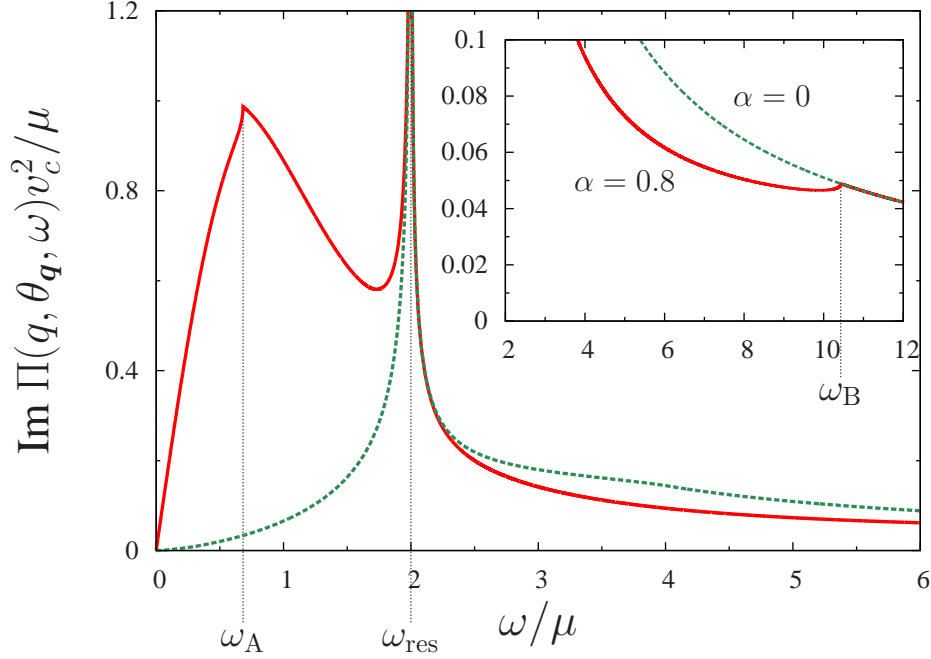}
\end{center}
\caption{
(Color online) 
Normalized imaginary part,
 ${\rm Im} \Pi(\theta_{\bf q},\omega) v_c^2/\mu$,  
   as a function of $\omega/\mu$
      with $q=2 \mu/v_c$ and $\theta_q=\pi/2$. 
 This behavior is expected in the case of $1/(1+\alpha \cos \theta_{\bm{q}}) < v_cq/\mu < 2/(\sqrt{1-\alpha^2} \sqrt{1 - \alpha^2 \cos^2 \theta_{\bm{q}} }) $.
The region of 
  $0 <\omega < \omega_{\rm res}$ 
  $(\omega > \omega_{\rm res} )$ 
 comes from  the intraband (interband) excitation. 
The dotted line represents the isotropic case.  
The cusp in the intraband region is found 
at $\omega = \omega_A$,  while 
 that in the interband region is found at $\omega = \omega_B$ (the inset).
}
\label{ImP_omg-dep_inter}
\end{figure}

The imaginary part exhibits a variety of $\omega$ dependences 
 depending on the magnitude of $\bm{q}$, which is divided into the 
following three regions.
In the region of the 
small $q$,
 a gap exists in the interband excitation, while  in the region of the large $q$ a gap exists in the intraband excitation. 
The intermediate region of $q$ is located  between
these two regions.

First, we show a novel  behavior seen for the intermediate magnitude of $\bm{q}$,
 which  comes from the tilted Dirac cone. 
In Fig. \ref{ImP_omg-dep_inter},  ${\rm Im} \Pi(\theta_{\bf q},\omega) v_c^2/\mu$ with fixed $v_cq/\mu = 2$ and $\theta_{\bm{q}} = \pi/2$ is shown   
as a function of $\omega/\mu $. There are two cusps at $\omega = \omega_A$  in the intraband region ($ 0 <\omega < \omega_{\rm res}$) 
and at $\omega = \omega_B$ in the interband region ($\omega >\omega_{\rm res} $), where $\omega_A$ and $\omega_B$ are given by eq.~(\ref{eq:omega_AB}). 
Note that  
\begin{align}
\label{eq:condition}
\omega_{\rm A} <\omega_{\rm res}<\omega_{\rm B} , 
\end{align}
since regions for intraband and interband excitations are separated by $\omega_{\rm res}$. 

\begin{figure}[tb]
\begin{center}
\includegraphics{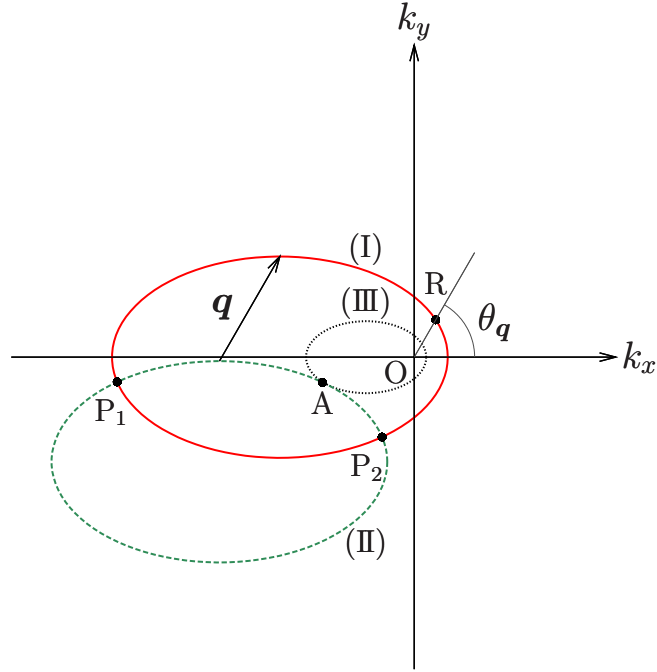}
\end{center}
\caption{
(Color online) 
Fermi surfaces for $\xi_{\bm{k}} = \mu$ (I),   $\xi_{\bm{k}+\bm{q}} = \mu$ $({\rm I\hspace{-0.1em}I})$,
 and  $\xi_{\bm{k}} = \mu - \omega_A$ $({\rm I\hspace{-0.1em}I\hspace{-0.1em}I})$
  on the $k_x$-$k_y$ plane. Point A where I and II contact gives rise to the emergence of the cusp. 
$P_1$ and $P_2$ are the intersection points between ellipse (I) and ellipse (II), and  O is the origin. The condition for the resonance 
 is obtained for  $k_y = k_x \tan \theta_{\bm{q}}$, which is found on the line 
 connecting O and R.   
}
\label{fig:fermisurface3}
\end{figure}

Here, using Fig. \ref{fig:fermisurface3}, we explain the cusp at $\omega = \omega_A$ in the intraband region.
Equation (\ref{eq:Polization}) shows  the following three kinds of conditions 
 for obtaining the imaginary part.  
 The first one is  the energy conservation given by 
\begin{align}
\omega =  \xi_{+, {{\bm k}+{\bm q}}} - \xi_{+, {\bm k}},
\label{eq:eh_excitation_energy}
\end{align}  
 and the second one is the possible process of the electron-hole excitation given by $f(\xi_{+,{\bm k}})-f(\xi_{+,{\bm k}+{\bm q}}) = 1$,
which leads to 
\begin{align}
\xi_{+,{\bm k}} < \mu , 
\label{xi_l_mu}
\\
\xi_{+,{\bm k}+{\bm q}} > \mu 
\label{xi_g_mu} .
\end{align}
The third one, which is discussed later is  a factor given by 
eq.~(\ref{eq:FF2}). 
 From the second condition,  it is found that  $\bm{k}$ is allowed inside of 
 ellipse (I) and outside of  ellipse (II) on the $k_x$-$k_y$ plane
 where ellipse (I) and ellipse (II) denote  
 $\xi_{+,{\bm k}} = \mu$ and $\xi_{+,{\bm k}+{\bm q}} = \mu$, respectively. 
By defining R as the intersection point between ellipse (I) and 
  $k_y = k_x \tan \theta_{\bm q }(> 0)$,
one finds $ \omega = \omega_{\rm res}$ being  the maximum 
 on the line given by $k_y = k_x \tan \theta_{\bm q} (> 0)$
(i.e., on the line OR).
 The cusp is understood as follows when  
the origin  O (i.e., $k_x=0$ and $k_y=0$ ) 
 is located outside  ellipse (II).  
We define $P_1$ and $P_2$ as the intersection points between ellipse (I) and ellipse (II) at which  $\omega (= \xi_{+, {\bm k}} -  \xi_{+, {{\bm k}+{\bm q}}})
 \rightarrow 0 $. 
Thus, $\omega$ increases from zero to $\omega_{\rm res}$
 when the point moves from $P_1$( or $P_2$) to R on ellipse (I).
Here, we consider ellipse (III), which 
has the same focus as  ellipse (I) and 
 touches  ellipse (II) at the point A.
It can be shown that such  ellipse (III) is given by 
\begin{align}
 \xi_{+,{\bm k}} = \mu - \omega_A ,
\label{eq:xi_A}
\end{align}
 where $\omega_A (>0)$ 
 is equal to that found in Fig. \ref{ImP_omg-dep_inter}. 
By noting  $\xi_{+,{\bm k}+{\bm q}} - \mu >0$ at $\bm{k} = 0$, and 
  the existence of $P_1$ and $P_2$,   
the condition for the cusp at $\omega_A$ is given by    
\begin{align}
 \frac{1}{1 + \alpha \cos \theta_{\bm q}}
 <  v_cq/\mu < \frac{2}{\sqrt{(1-\alpha^2)(1-\alpha^2 \cos^2 \theta_{\bm q})}}  ,
\label{eq:q_condition_lower}
\end{align}
 where $\omega_{\rm res} > \omega_A  > 0$.
On ellipse (II), $\omega$ takes a minimum at $P_1$ (and  $P_2$)
 and a maximum at A
 suggesting a  saddle point at A, which 
 gives rise to the cusp.  
Actually, 
this leads to an excess  contribution for the imaginary part, 
 which becomes singular for $\omega \rightarrow \omega_A - 0$, i.e.,  
\begin{align}
 \delta \left( {\rm Im} \Pi(q, \theta_{\bm q},\omega) \right) 
 \propto - \sqrt{\omega_A - \omega}. 
\label{eq:singular}
\end{align}

\begin{figure}[tb]
\begin{center}
\includegraphics{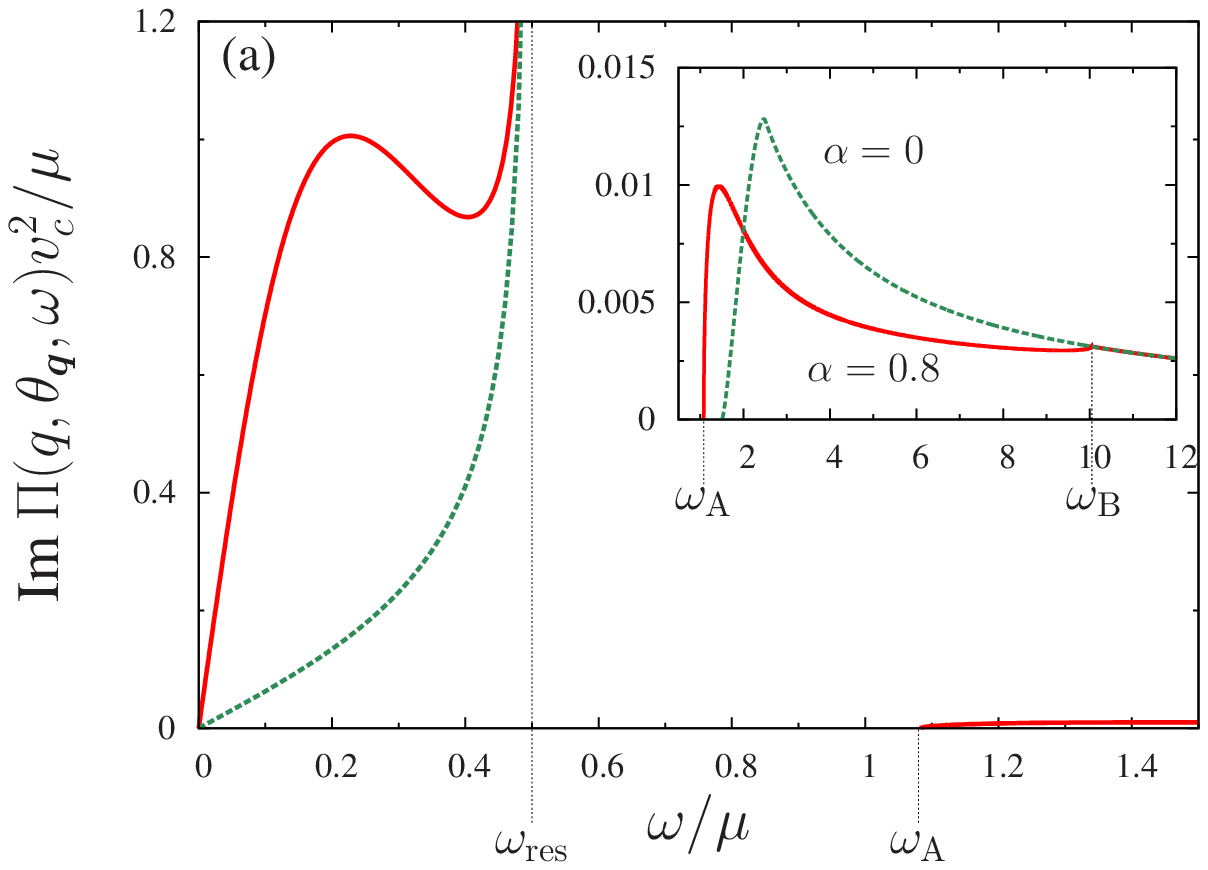}
\includegraphics{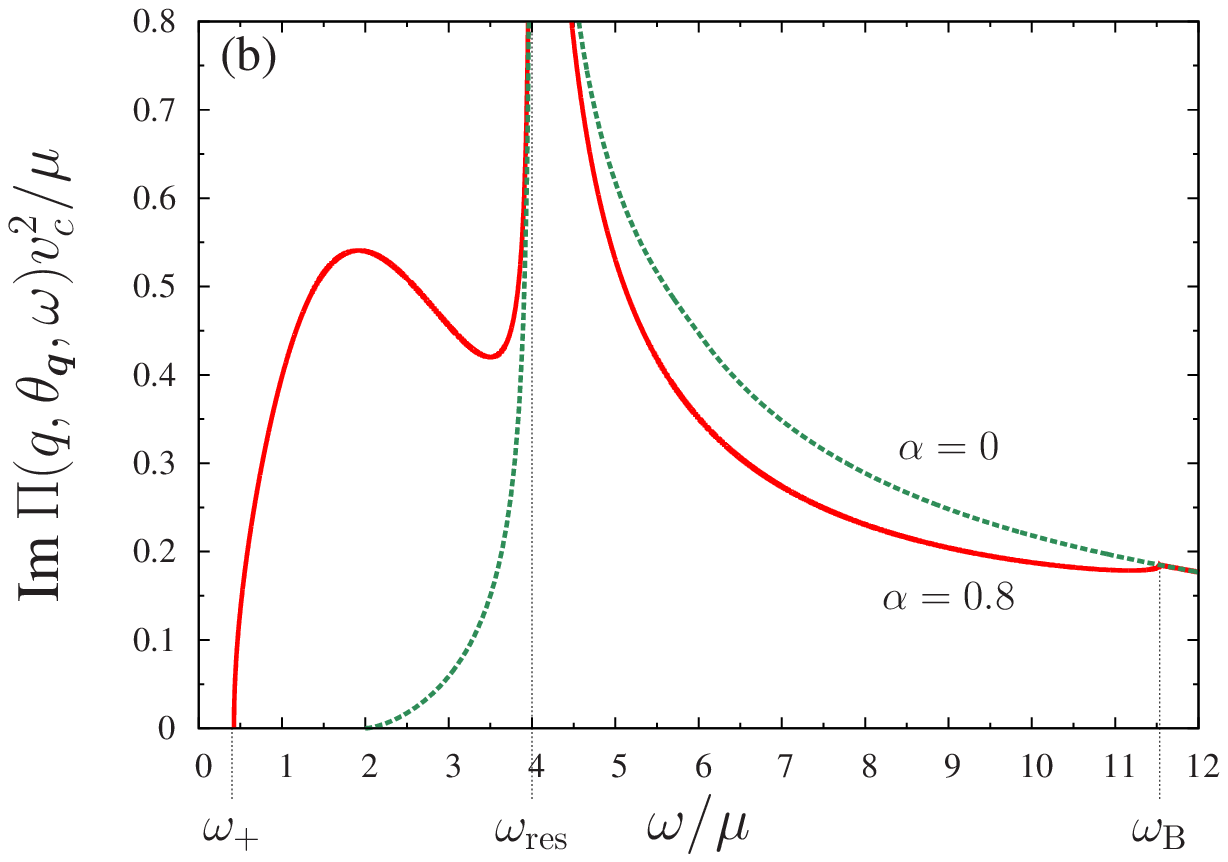}
\end{center}
\caption{
(Color online) 
Normalized imaginary part,
 ${\rm Im} \Pi(\theta_{\bf q},\omega) v_c^2/\mu$,  
 as a function of $\omega/\mu$
  in the case of  $\theta_q=\pi/2$
   for $q= \mu/2v_c$ (a) (upper panel)
   and for $q= 4 \mu/v_c$ (b) (lower panel). 
The behaviors of panels (a) and (b) are found  
  for  $  v_cq/\mu < 1/(1+\alpha \cos \theta_{\bm{q}})$ 
 and $ 2/(\sqrt{1-\alpha^2} \sqrt{1 - \alpha^2 \cos^2 \theta_{\bm{q}} })< v_cq/\mu$, respectively.
The result in the isotropic case is shown by the dashed curve.
}
\label{Figure:fig5}
\end{figure}
\begin{figure}[tb]
\begin{center}
\includegraphics{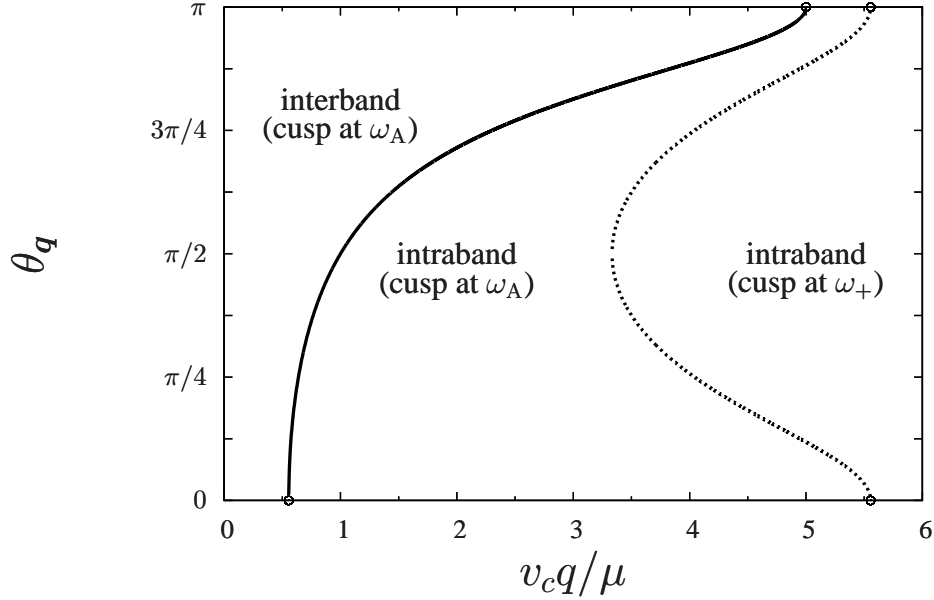}
\end{center}
\caption{
Region for several cusps on the $v_cq/\mu$ - $\theta_{\bm{q}}$ plane, 
where $\omega_{\rm A}$ is found for the intermediate  $v_cq/\mu$ 
with  the fixed
 $\theta_{\bm{q}}$ (eq.~(\ref{eq:q_condition_lower})). 
}
\label{fig:cusp}
\end{figure}


Next we examine the imaginary part in the other cases
 of the small $q$ and the large $q$, which are found for 
 $  v_cq/\mu < 1/(1+\alpha \cos \theta_{\bm{q}})$  and $ 2/(\sqrt{1-\alpha^2} \sqrt{1 - \alpha^2 \cos^2 \theta_{\bm{q}} })< v_cq/\mu$, respectively.
These examples are shown in Figs. \ref{Figure:fig5}(a) and 5(b) by choosing 
   $q= \mu/2v_c$ (a) (upper panel) and  $q= 4 \mu/v_c$ (b) (lower panel).
Note that the cusp is also found at $\omega_{A}$ or $\omega_{+}$, 
which can be understood from the saddle point similar to 
 that in  Fig.~\ref{fig:fermisurface3}. 
 There is a broad peak for $\omega < \omega_{\rm res}$, which is also characteristic of a tilted cone and is absent in the isotropic case. 
The location of such a peak is almost proportional to $q$. 
A noticeable difference  is seen between the small $q$ and the large $q$ 
    (i.e., between the case  $q= \mu/2v_c$ and that of $q= 4 \mu/v_c$).
  For $q= \mu/2v_c$ (upper panel), 
  the interband contribution of ${\rm Im} \Pi(\theta_{\bf q},\omega) v_c^2/\mu$      is very small and 
      increases from zero  with increasing $\omega (> \omega_{\rm A})$.  
 As seen from Figs. \ref{ImP_omg-dep_inter} and \ref{Figure:fig5} (a), ${\rm Im} \Pi(\theta_{\bf q},\omega)$ of the intraband excitation 
    ($\omega < \omega_{\rm res}$) is much larger than
 that of the interband excitation ($\omega > \omega_{\rm res}$).
An opposite behavior emerges 
 when $v_c q/\mu >> 1$ (Fig. \ref{Figure:fig5} (b)). 
For $q= 4 \mu/v_c$ (lower panel), the interband contribution of ${\rm Im} \Pi(\theta_{\bm q},\omega) v_c^2/\mu$ is large 
 and is close to $ \omega \sim \omega_{\rm res}$. 
Compared with that in the isotropic case, the intraband contribution is enhanced but the interband contribution is suppressed.
Such a fact may be  ascribed to a sum rule given by 
$
\int_{0}^{\Lambda} {\rm d} \omega \omega {\rm Im} \Pi (\bm{q},\omega)
 = (1/8) q^2 \Lambda
$, 
 where $\Lambda$ is the upper cutoff of the frequency. 
\cite{Sabio2008}
For $ v_c q /\mu <  1/(1+\alpha \cos \theta_{\bm q}) $,  where 
 $\omega_A >  \omega_{\rm res}$,  the imaginary part   is absent  in the interval range of $\omega_{\rm res} < \omega < \omega_A$.
For a small $\omega$, the imaginary part is convex downward in the isotropic case  owing  to the absence of  the backward scattering, 
 while it is convex upward  in the tilted case owing  to 
  $P_1$ and $P_2$ located away from the backward scattering.    
Note that the cusp at $\omega_B$ is present for the arbitrary $q$, 
    as seen from Fig. \ref{ImP_omg-dep_inter}.

 We examine the $\theta_{\bf{q}}$  dependence  of the imaginary part. 
 On the basis of  eq.~(\ref{eq:q_condition_lower}),  the region for the existence of 
$\omega_A$ is shown on the $v_cq/\mu$-$\theta_{q}$ 
plane in Fig. \ref{fig:cusp}. 
 The behavior for  the fixed $\theta_{\bf{q}}(= \pi/2)$ is as follows. 
In the region for the small $v_cq/\mu$, 
the cusp at $\omega = \omega_A$ appears in the interband process 
 (Fig. \ref{Figure:fig5}(a)), 
 while in the middle region with an intermediate $v_cq/\mu$, $\omega_A$ exists  in  the intraband process (Fig. \ref{ImP_omg-dep_inter}).
 For the large  $v_cq/\mu$, the cusp appears at $\omega_{+}$,  which is the lower bound of the imaginary part (Fig. \ref{Figure:fig5} (b)).
Note that the cusp is found, 
  except for  $\theta_{\bm{q}} = 0$ and $\pi$.

Here, we briefly mention  the $\theta_{\bf q}$ dependences on  $\omega_{\rm A}$, $\omega_{\rm res}$, and $\omega_{\rm B}$ with a fixed $v_cq/\mu$.  
 With increasing $\theta_q$ from zero to $\pi$,
  both $\omega_{\rm B}$ and $\omega_{\rm res}$ decrease monotonically, 
 but $\omega_{\rm A}$ remains
  almost constant until  $\omega_A = \omega_{\rm res}$.
The cusp at $\omega_{\rm A}$ moves from the intraband region to the interband region, where the boundary is given by  $ \omega_{\rm res}$. 
Those frequencies also depend on the degree of tilting, $\alpha$.
With increasing $\alpha$ from zero to 1,  both $\omega_{\rm A}$ and  $\omega_{\rm B}$ increase monotonically. The resonance frequency $\omega_{\rm res}$ increases (decreases)
 for $0< \theta_{\bf q}< \pi/2$ ($ \pi/2 < \theta_{\bf q}< \pi)$
  with increasing  $\alpha$, 
  while  $\omega_{\rm res}$ at $\theta_{\bf q} = \pi/2$ 
   is independent of $\alpha$.   

\begin{figure}[tb]
\begin{center}
\includegraphics[width=0.5\textwidth,height=0.5\textwidth,keepaspectratio=true]{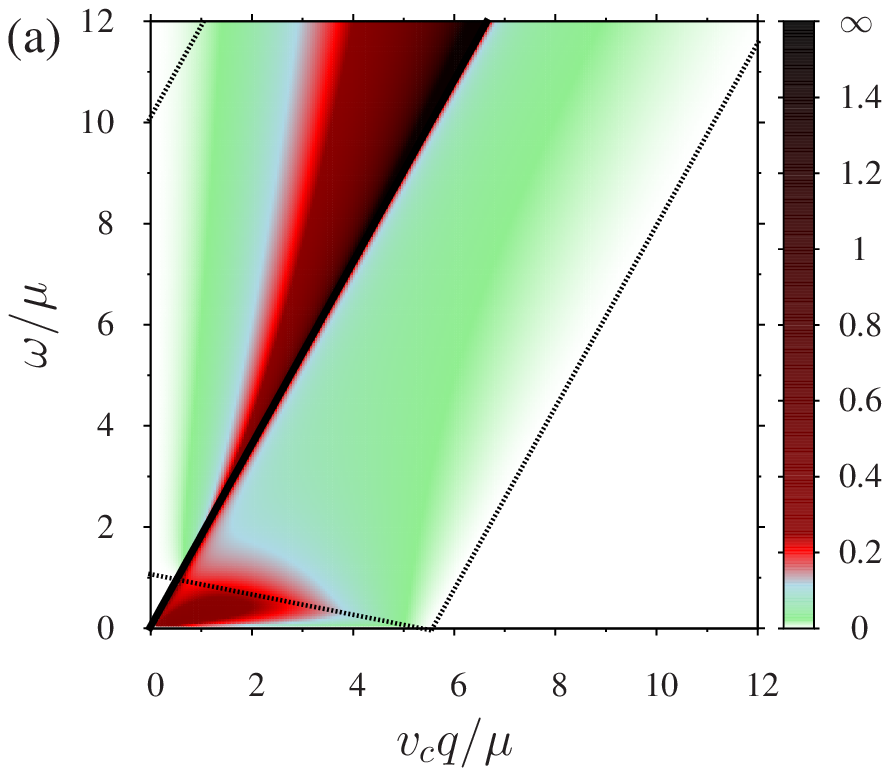}
\includegraphics[width=0.5\textwidth,height=0.5\textwidth,keepaspectratio=true]{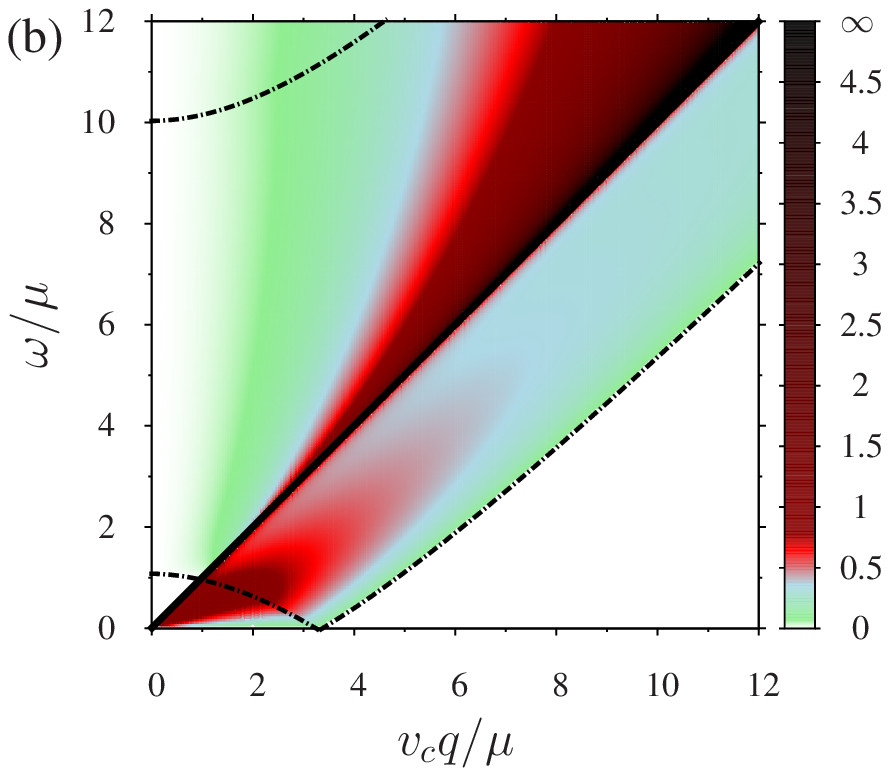}
\includegraphics[width=0.5\textwidth,height=0.5\textwidth,keepaspectratio=true]{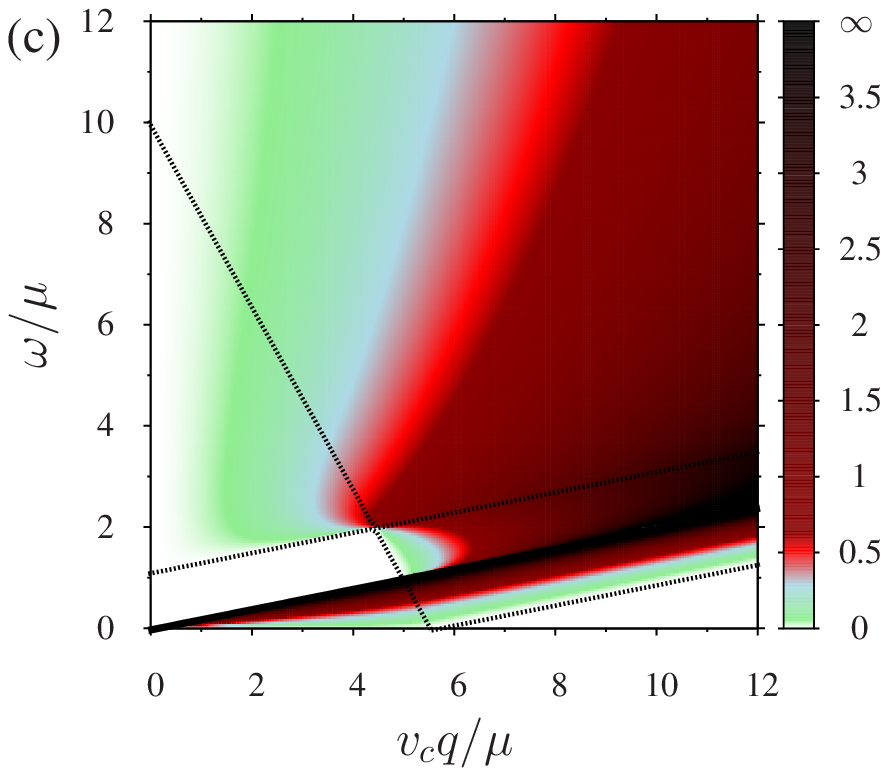}
\end{center}
\caption{
Normalized imaginary part,
 ${\rm Im} \Pi(\theta_{\bf q},\omega) v_c^2/g_s\mu$ 
 on the plane of $v_c q/\mu$ and $\omega / \mu$
 for  $\theta_q= 0$ (a),  $\theta_q=\pi/2$ (b), 
 and $\theta_q=\pi$ (c), where $\alpha=0.8$.
The straight line including the origin represents the resonance frequency $\omega_{\rm res}$.
The dark (bright) area represents the large (small)   
${\rm Im} \Pi(\theta_{\bf q},\omega)$, while 
${\rm Im} \Pi(\theta_{\bf q},\omega)=0 $ in 
the white area corresponding to the  regions of 3A and 1B in Fig. \ref{ImP_region_2pi4}.
The dash-dotted line (dotted line) denotes the boundary 
 for the excitation process at which the cusp emerges (is absent). 
}
\label{ImP_q-omg_0pi4}
\end{figure}

Figure \ref{ImP_q-omg_0pi4} shows the normalized imaginary part 
 ${\rm Im} \Pi(\theta_{\bf q},\omega) v_c^2/\mu$ on the plane of $v_c q/\mu$ and $\omega / \mu$ for $\theta_{\bm{q}} =$ 0(a), $\pi/2$(b), and  $\pi$ (c).
The color gauge with the gradation represents the magnitude of the imaginary part. The global feature is mainly determined by the property of $\omega_{\rm res}$.  In the case of $v_c q/\mu >> 1$ where the characteristic energy becomes much larger than the interband energy ($\sim \mu$), 
${\rm Im} \Pi(\theta_{\bf q},\omega)$ of the intraband excitation ($\omega < \omega_{\rm res}$) becomes much  smaller than that of the interband excitation ($\omega > \omega_{\rm res}$) in contrast to the case  where  $v_c q/\mu = 2$  
(Fig. \ref{ImP_omg-dep_inter}).
 The broad peak in the intraband excitation 
(i.e., $\omega < \omega_{\rm res}$) 
  does not change much for $\theta_{\bm{q}} < \pi/2$, while it is strongly  masked for $\pi/2 < \theta_{\bm{q}} < \pi$
due to the rapid decreasing  $\omega_{\rm res}$. 
The  cusp at $\omega=\omega_A$ is seen for $0 < \theta_{\bm{q}} < \pi$, 
 as seen also from the analytical condition 
of eq. (\ref{eq:q_condition_lower}).
The cusp in the interband process 
 does exist in an intermediate region of $q$ 
 for the arbitrary $\theta_{\bm q}$ (except for 0 and $\pi$).

\begin{figure}[tb]
\begin{center}
\includegraphics[width=0.5\textwidth,height=0.5\textwidth,keepaspectratio=true]{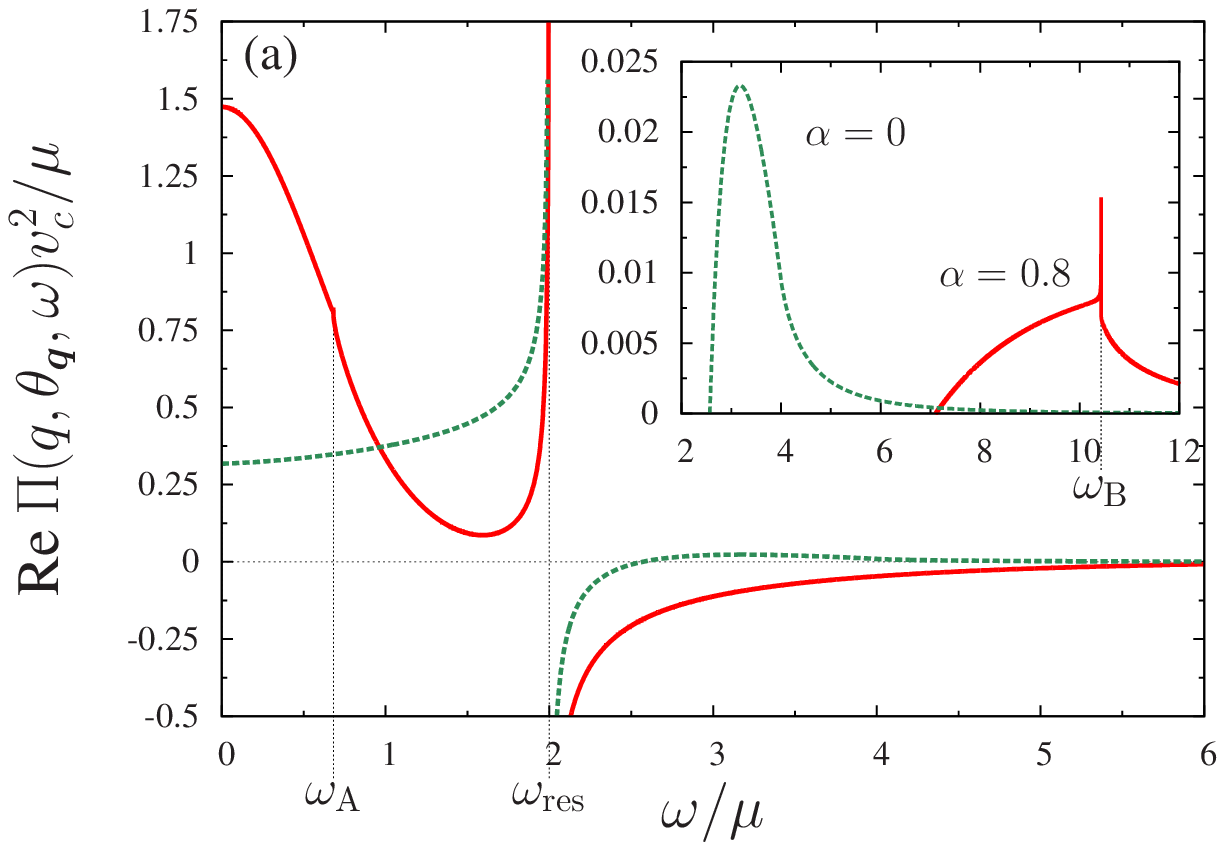}
\includegraphics[width=0.5\textwidth,height=0.5\textwidth,keepaspectratio=true]{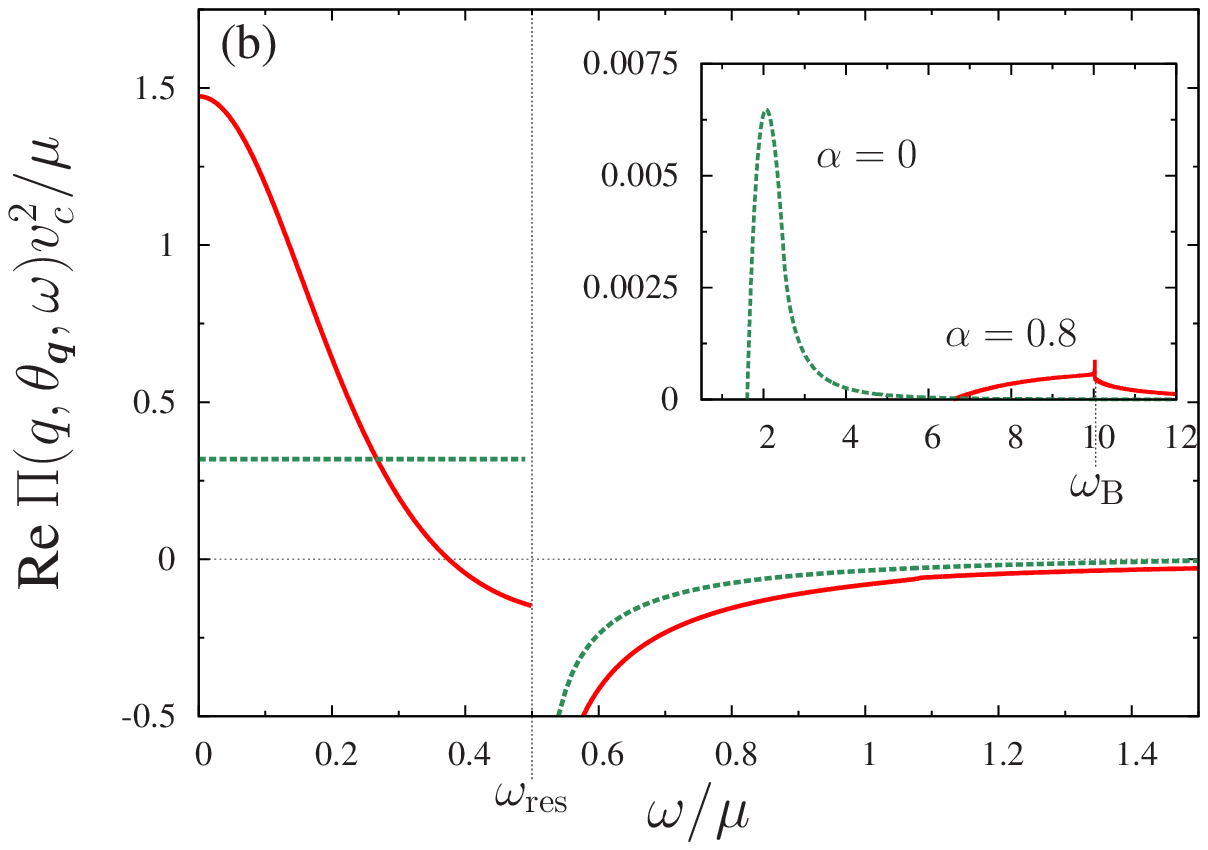}
\includegraphics[width=0.5\textwidth,height=0.5\textwidth,keepaspectratio=true]{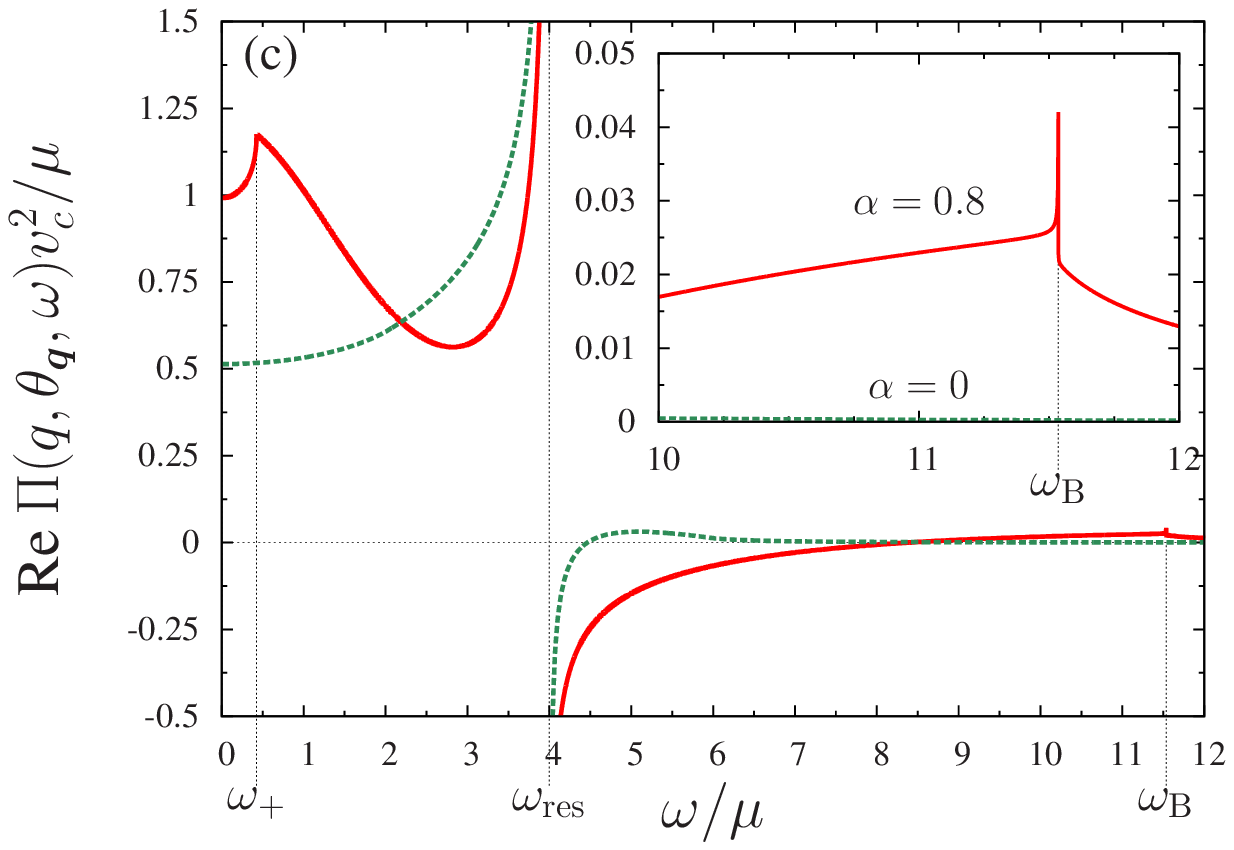}
\end{center}
\caption{
(Color online) 
Real part Re$\Pi (q, \theta_{\bm q}, \omega )$ as a function of $\omega/\mu$
with a fixed $\theta_{\bm q} = \pi/2$, 
 for $q v_c/\mu$ = 2 (a), 1/2(b), and 4(c),   which correspond to 
Figs.~3 and 5(a) and 5(b). 
The global feature with arbitrary $q$ is seen 
 for $\omega > \omega_{\rm res}$.
The divergence of the real part for $\omega \rightarrow \omega_{\rm res} -0$
 is obtained for $q > 1/(1 + \alpha \cos \theta_q)$.
The dotted line represents the isotropic case of graphene. 
}
\label{ReP_omg}
\end{figure}

Finally in this subsection, we examine the real part, which is calculated using eq.~(\ref{eq:KK}) (Appendix C). The numerical results of 
${\rm Re} \Pi(q,\theta_{\bm{q}},\omega) v_c^2/\mu$ corresponding to the imaginary part of Figs.~\ref{ImP_omg-dep_inter}, \ref{Figure:fig5}(a),  and 
 \ref{Figure:fig5}(b) are shown in Fig.~\ref{ReP_omg}. 
  When $q > (\mu/v_c)/(1+\alpha \cos \theta_{\bm q})$
 as seen from  Figs. \ref{ReP_omg}(a) and \ref{ReP_omg}(c), 
${\rm Re} \Pi (q, \theta_{\bm q}, \omega)$  diverges for $\omega \rightarrow  \omega_{\rm res} - 0$.  The cusp found in the imaginary part also appears in the real part, e.g., $\omega_{\rm A}$ and $\omega_{\rm B}$ for $ v_cq/\mu =2$,  $\omega_{\rm B}$ for $ v_cq/\mu =1/2$, and $\omega_+$ and $\omega_{\rm B}$ for $ v_cq/\mu =4$. A noticeable difference compared with the isotropic case is the nonmonotonic behavior for $\omega < \omega_{\rm res}$ and $v_cq/\mu > 2/(1+\alpha \cos \theta_{\bm{q}})$, e.g., exhibiting a minimum  for $v_cq/\mu =$ 2 and 4.
 In the regime of the interband  ( $\omega > \omega_{\rm res}$),
the real part is similar but is slightly small compared with that in 
 the isotropic case.

\section{Optical Conductivity, Plasma Mode, and Screening Effect}
\subsection{Optical conductivity}
The optical conductivity is obtained from  two kinds of 
electron-hole excitations, i.e., the intraband processes 
 and interband processes.
 The intraband excitation  gives a Drude-like behavior in the presence of 
the impurity scattering, while  the interband excitation  gives 
a gaplike behavior.
In this subsection, we calculate the interband optical conductivity, 
while the effect of  the intraband one is left for discussion. 

\begin{figure}[tb]
\begin{center}
\includegraphics{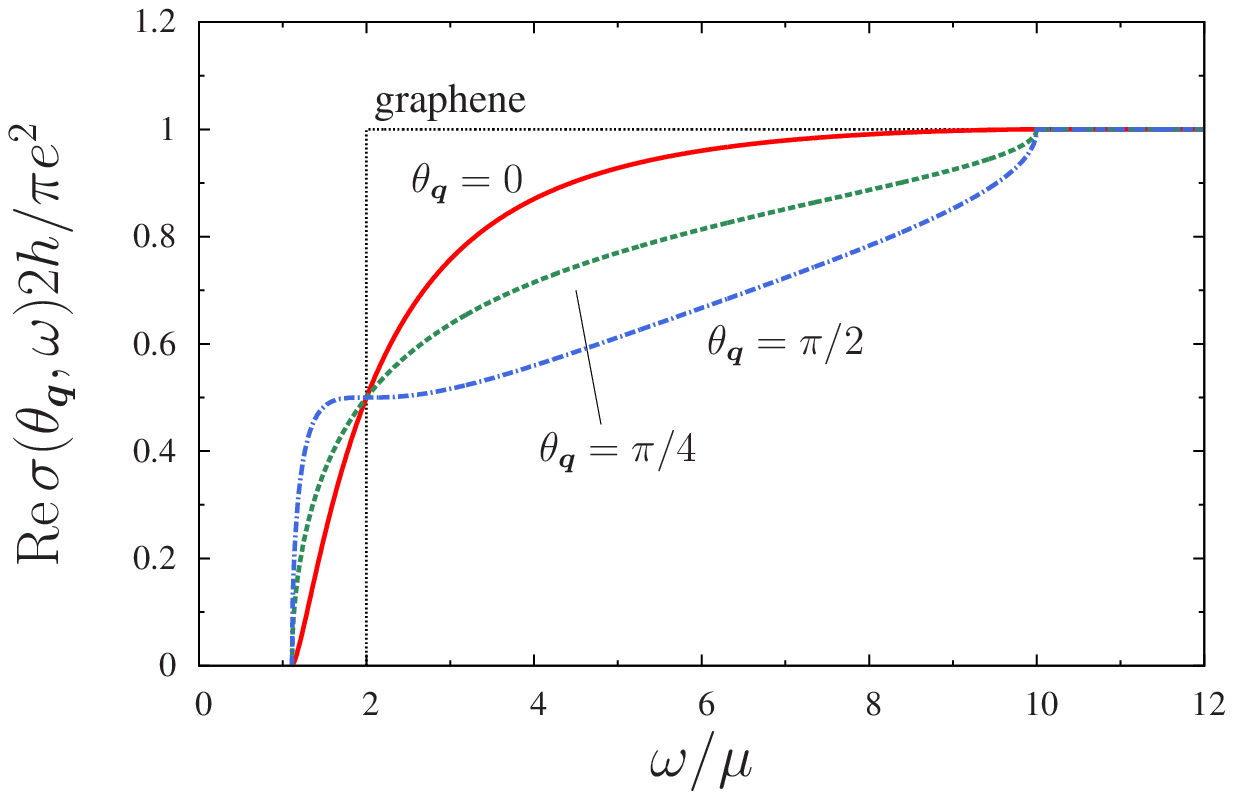}
\end{center}
\caption{
(Color online) 
Normalized interband optical conductivity ${\rm Re} \sigma(\theta_q,\omega)2 h/\pi e^2$ as a function of $\omega/\mu $
 for $\theta_{\bm q}$ = 0, $\pi/4$ and $\pi/2$. 
The contribution comes from only the  interband excitation. 
Note that ${\rm Re} \sigma (\theta_{\bm q}) = {\rm Re} \sigma ( \pi - \theta_{\bm q}).$
The dotted line followed by a jump represents the isotropic case of graphene.
}
\label{ReSgm_omg-dep}
\end{figure}

\begin{figure}[tb]
\begin{center}
\includegraphics{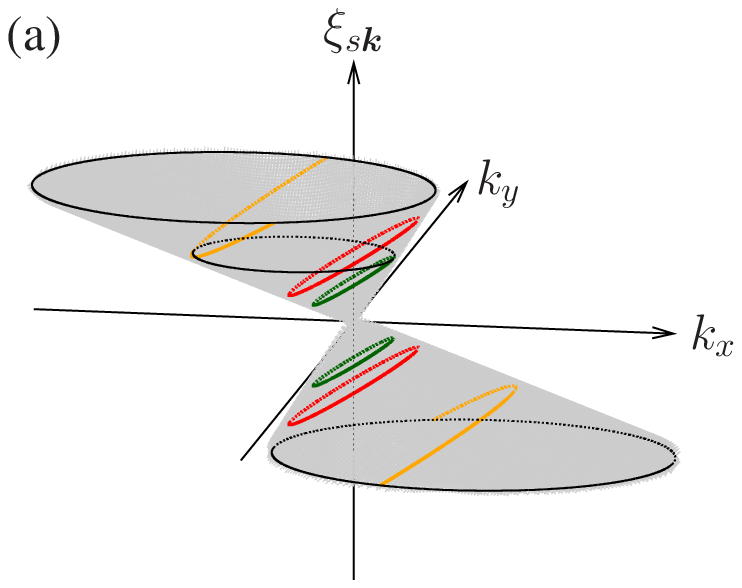}
\includegraphics[width=0.5\textwidth,height=0.5\textwidth,keepaspectratio=true]{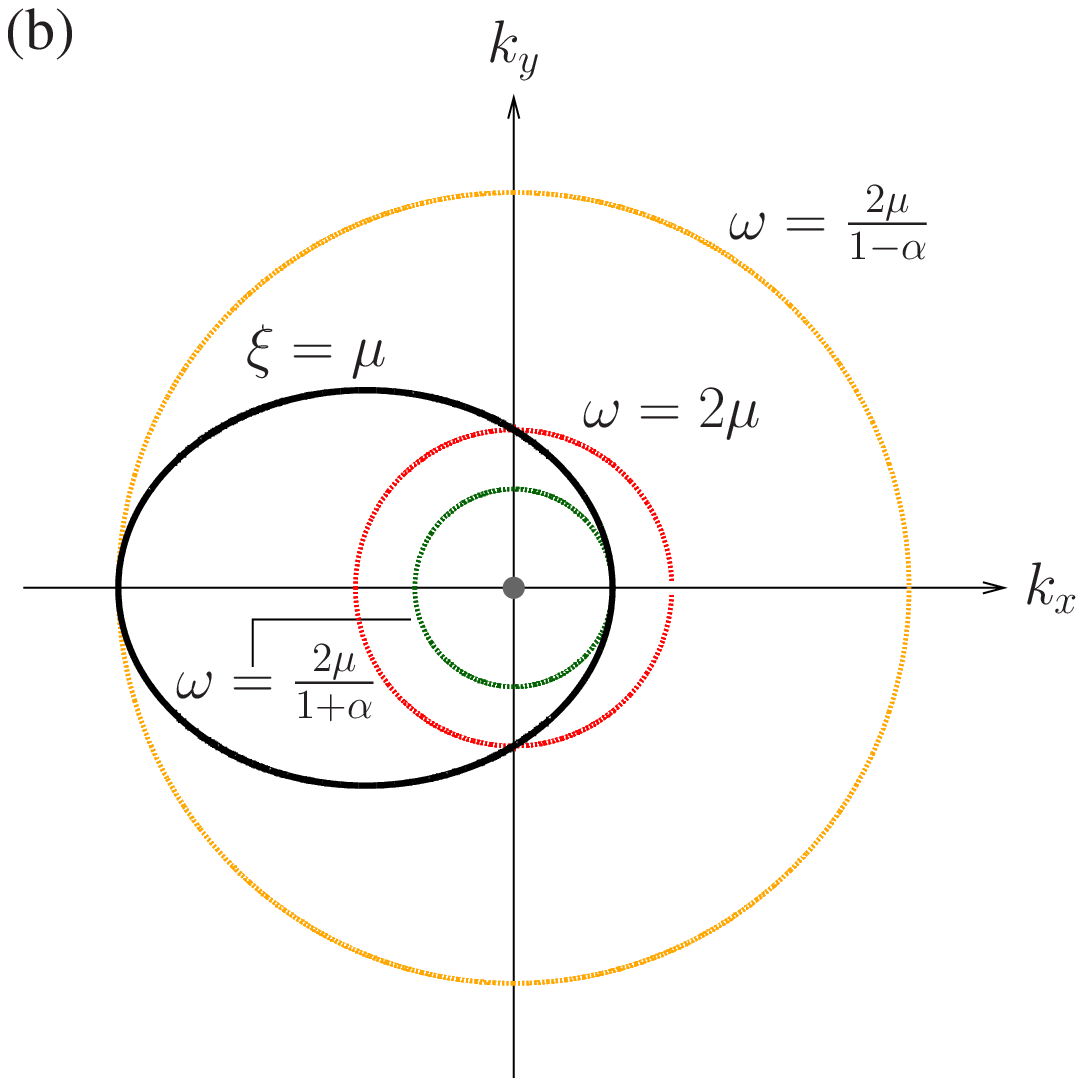}
\end{center}
\caption{
(Color online) 
(a) Particle-hole excitation with zero momentum transfer 
 from the lower cone to the upper cone where three kinds of ellipses 
 lead to the energy with  
   $\omega = 2 \mu /(1 + \alpha)$, 2$\mu$ 
   and  2 $\mu/(1 - \alpha)$. 
(b) The three kinds of ellipses of (a) are projected 
 on the  $k_x$-$k_y$ plane. The excitation is  allowed 
  only in the region which is located 
    outside the  Fermi surface (solid ellipse).
}
\label{opt}
\end{figure}

In terms of the imaginary part of $\Pi({\bm q},\omega)$, 
 the optical conductivity $\sigma (\omega)$ is calculated as
\cite{Pines1966}
\begin{align}
\sigma(\omega)=
\lim_{q\rightarrow 0}
\frac{ie^2\omega}{\hbar q^2}\left\{\Pi({\bm q},\omega)+\Pi(-{\bm q},\omega)\right\}, 
\end{align}
where   the RPA gives the same result. 
Since  ${\rm Im}\,\Pi (0,\omega)$ 
 becomes finite  only for the interband electron-hole excitation,
 ${\rm Re}\,\sigma(\omega)$ is calculated as
 (Appendix \ref{sec:cal_Im})
\begin{align}
{\rm Re}\,\sigma(\omega)=
&0\cdot\Theta\left(\frac{2\mu}{1+\alpha}-\omega\right)
\nonumber\\
+&\frac{e^2}{8\pi\hbar}
\Biggl[
\pi
+{\rm sgn}\left(\frac{2\mu}{1+\alpha|\cos\theta_{\bm q}|}-\omega\right)
G_{<}(z_2^{+})
+{\rm sgn}\left(\frac{2\mu}{1-\alpha|\cos\theta_{\bm q}|}-\omega\right)G_{<}(-z_2^{-})
\Biggr]
\nonumber\\
&\times
\Theta\left(\frac{2\mu}{1-\alpha}-\omega\right)
\Theta\left(\omega-\frac{2\mu}{1+\alpha}\right)
\nonumber\\
+&\frac{e^2}{8\pi\hbar}
2\pi
\Theta\left(\omega-\frac{2\mu}{1-\alpha}\right),
\label{eq:sgm}
\end{align}
where 
$\Theta$ is a step function, ¡¡
$G_{<}(x)=x\sqrt{1-x^{2}}-\arccos(x)\quad{\rm for}\quad |x|<1$, 
and  $z_2^{\pm}$ 
are  defined by
\begin{align}
&z_2^{\pm}=\frac{|\cos\theta_{\bm q}|}{\alpha}\left(\frac{2\mu}{\omega}-1\right)
\pm\frac{|\sin\theta_{\bm q}|}{\alpha}\sqrt{\alpha^2-\left(\frac{2\mu}{\omega}-1\right)^{\!\!2}} . 
\end{align}
In Fig.~\ref{ReSgm_omg-dep}, the normalized optical conductivity
${\rm Re} \sigma(\theta_q,\omega)4 \hbar/e^2$ is shown 
 as a function of $\omega/\mu $ for $\theta_{\bm q} =0$, $\pi/4$ and 
 $\pi/2$, respectively,  with  $\alpha =0.8$. 
 The dotted line corresponds to the isotropic case  where the normalized ${\rm Re} \sigma (\theta_{\bm q},\omega)$ is zero for $\omega < \mu$ and is constant for $\mu < \omega$. 
The conductivity ${\rm Re} \sigma (\theta_{\bm q},\omega)$ is zero 
 for $\omega <2 \mu/(1+\alpha )$ 
 and 1  for $\omega > 2\mu /(1 - \alpha)$, while 
it takes an intermediate value for  $2 \mu/(1+\alpha ) < \omega <  2\mu /(1-\alpha )$. 
The reason for such a behavior is illustrated in Figs.~\ref{opt}(a)
 and \ref{opt}(b). 
The electron-hole excitation for the conductivity is a vertical transition from the valence band to the conduction band due to $q=0$. 
Figure~\ref{opt}(a) denotes the process with 
 a fixed energy $\omega = \xi_{+ \bm{k}} - \xi_{- \bm{k}}$.
The contour projected on the $k_x$-$k_y$ plane 
is shown by a circle in  Fig.~\ref{opt}(b). 
The ellipse of the thick line denotes the Fermi surface with $\xi=\mu$. The electron-hole excitation is allowed outside of the thick line since the hole is created above the Fermi surface of the conduction band. 
 Thus, the process is completely forbidden for $\omega <2 \mu/(1+\alpha )$,   while such a process occurs on all the corresponding circles for $\omega > 2\mu /(1 - \alpha)$. 
In the intermediate case of $2\mu /(1 + \alpha) < \omega < 2 \mu/(1-\alpha )$, 
the electron-hole excitation occurs on the partial part of the circles outside the thick line, i.e., for $\theta_{\bm q}$ satisfying the condition $\omega > 2 \mu/(1+\alpha  \cos \theta_{\bm q} )$.  The tilting of the Dirac cone is essential for the optical conductivity  that  depends on $\theta_{\bm q}$ .

The conductivity can be calculated directly by taking the limit of $q \rightarrow 0$ in the imaginary part of the dielectric function. 
From eq.~(\ref{eq:Polization}),
 the polarization function, which includes  the excitation from the valence band
  to the  conduction band is written as 
\begin{align}
\label{eq:Plasma_1}
\Pi_{-+}({\bm q},\omega)
&
=-\frac{2}{(2\pi)^2}\int\hspace{-0.3em}{\rm d}{\bm k}  
\left\{
1-\frac{{\bm k}\!\cdot\!({\bm k}+{\bm q})}{|{\bm k}||{\bm k}+{\bm q}|}
\right\}
\frac{1-f(\xi_{+,\bm k})}
{\omega+i\eta+\xi_{{-,\bm k}-{\bm q}}-\xi_{+,\bm k}}.
\end{align}
By expanding up to $O$($q^2$), the imaginary part for the small $|{\bm q}|$ is rewritten as 
\begin{align}
\label{eq:Plasma_2}
{\rm Im}\,\Pi_{-+}({\bm q},\omega)
&
\rightarrow
\frac{4 \pi}{(2\pi)^2}\int\hspace{-0.3em}{\rm d}\varphi
\hspace{-0.3em}
\int\hspace{-0.3em}{\rm d}k\,k
\frac{1-\cos^2(\varphi-\theta_{\bm q})}{4v_c k^2}q^2
\nonumber\\
&\times
\delta\left(\frac{\omega}{v_c}-2k-\alpha q_x-q\cos(\varphi-\theta_{\bm q})\right)
\Theta\left(k+\alpha k\cos\varphi-\frac{\mu}{v_c}\right)
\nonumber\\
=
&
\frac{q^2}{\omega}\frac{\pi}{(2\pi)^2}
\int\hspace{-0.3em}{\rm d}\varphi
\left\{1-\cos^2(\varphi-\theta_{\bm q})\right\}
\Theta\left(\omega-\frac{2\mu}{1+\alpha\cos\varphi}\right)
\nonumber\\
\equiv
&
\frac{q^2}{4\omega}F_c(\theta_{\bm q}).
\end{align}
The quantity $F_c({\bm q})$ is zero for $\omega<2\mu/(1+\alpha)$ 
and $F_c({\bm q}) = 1$ for $2\mu/(1-\alpha)<\omega$, and 
\begin{align}
\label{eq:Plasma_3}
F_c(\theta_{\bm q})
=
&
\frac{1}{\pi}\int_{-\varphi_0}^{\varphi_0}\hspace{-1.3em}{\rm d}\varphi
\left\{1-\cos^2(\varphi-\theta_{\bm q})\right\} 
\nonumber\\
=
&
\frac{1}{\pi}
\left(\varphi_0-\frac{1}{2}\sin 2\varphi_0 \cos 2\theta_{\bm q}\right)
\end{align}
for $2\mu/(1 +\alpha)<\omega<2\mu/(1-\alpha)$, where 
\begin{align}
\label{eq:Plasma_4}
\cos\varphi_0=\frac{1}{\alpha}\left(\frac{2\mu}{\omega}-1\right).
\end{align}
Thus, we obtain the optical conductivity as 
\begin{align}
\label{eq:Plasma_5}
\sigma (\omega) = \frac{e^2 }{4 \hbar}  F_c(\theta_{\bm q}) . 
  \end{align}
One finds the following $\theta_{\bm q}$ dependence of the optical conductivity 
in the interval region of 
$2\mu/(1 +\alpha)<\omega<2\mu/(1-\alpha)$. 
From eq.~(\ref{eq:Plasma_3}), 
$\sigma$ with a fixed $\omega$ takes a maximum (minimum) at $\theta_{\bm q}=\pi/2$ 
and a minimum (maximum) at $\theta_{\bm q} = 0$ and $\pi$ 
for $\omega<2\mu\,(\omega>2\mu)$.

\subsection{Plasma mode}
\begin{figure}[tb]
\begin{center}
\includegraphics{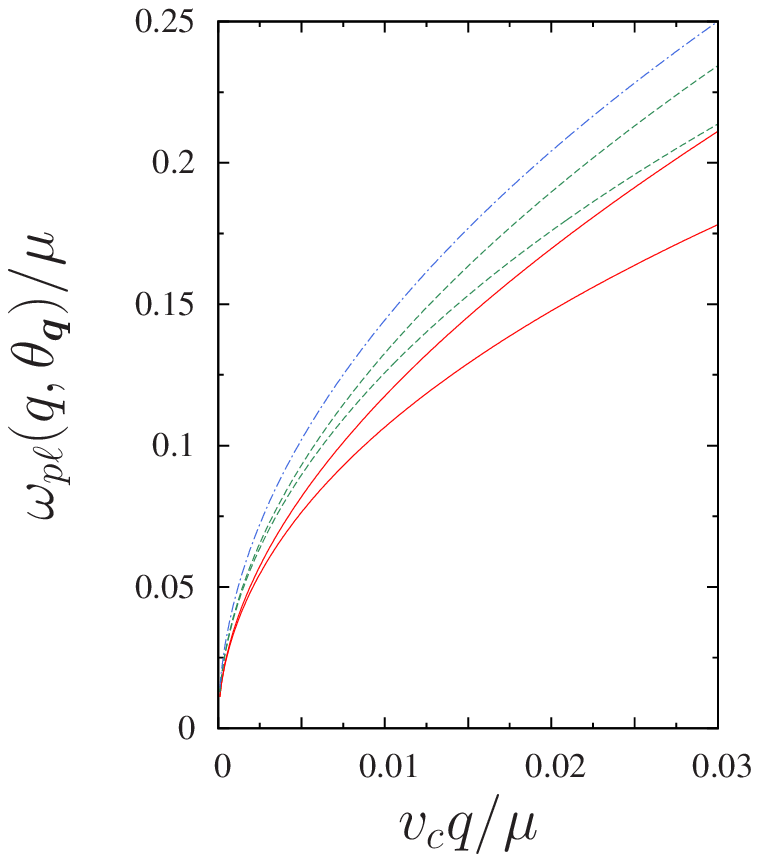}
\end{center}
\caption{
(Color online) 
Plasma frequency as a function of $v_c q /\mu$
 for $\theta_{\bm{q}} = \pi/2, \pi/4, 3\pi/4, 0, \pi$ 
(from the top to the bottom). 
}
\label{Plasmon}
\end{figure}
We examine the plasma frequency by adding the Coulomb interaction.
Within the RPA, the plasma frequency can be calculated from 
the pole of the polarization function, which is obtained by  
\begin{align}
1+v_{\bm q}{\rm Re} \,\Pi(q,\theta_{\bm q},\omega)=0 ,
 \label{eq:eq_RPA_denom}
\end{align}
where $v_{\bm q}$ is the Fourier transform of the Coulomb interaction,
\begin{align}
 v_{\bm q} 
    = \frac{2\pi e^2}
        {\epsilon_0 |{\bm q}|} \; .
\end{align}
Equation (\ref{eq:eq_RPA_denom}) is written explicitly as
\begin{align}
\label{eq:Collect_1}
1 & = -  v_{\bm q}  
\frac{2}{L^2}\sum_{{\bm k},s=+,s'=+}
\big|{\bm F}^{\dagger}_{s}({\bm k}){\bm F}_{s'}({\bm k}+{\bm q})\big|^2
\frac{f(\xi_{s{\bm k}})-f(\xi_{s'{\bm k}+{\bm q}})}
{\omega+i\eta-(\xi_{s'{\bm k}+{\bm q}}-\xi_{s{\bm k}})}
 \nonumber \\ 
&
=-v_{\bm q}\frac{1}{(2 \pi)^2}
\int_{0}^{\omega}\hspace{-0.8em}{\rm d}k\,k
\int_{0}^{2\pi}\hspace{-1em}{\rm d}\varphi f(\xi_{\bm k})
\nonumber\\ 
&
\times\left[
\left\{
1+ \frac{{\bm k}\!\cdot\!({\bm k}+{\bm q})}{|{\bm k}| | {\bm k}+{\bm q}|}
\right\}
\frac{1}{\omega + \xi_{\bm k}- \xi_{{\bm k} + {\bm q}}}
-\left\{
1+ \frac{{\bm k}\!\cdot\!({\bm k}-{\bm q})}{|{\bm k}| | {\bm k}-{\bm q}|}
\right\}
\frac{1}{\omega - \xi_{\bm k}+ \xi_{{\bm k} + {\bm q}}}
\right]
\; , 
\end{align}
 which is calculated using the real part of $\Pi(q,\theta_{\bm q},\omega)$ obtained in \S 3.
In Fig. \ref{Plasmon}, plasma frequency as a function of $v_c q /\mu$ is shown with several choices of  $\theta_{\bm{q}}$. 
In order to understand the behavior with a small $q$,  
 we calculate  the collective mode by expanding eq.~(\ref{eq:Collect_1})
 in terms of ${\bm q}$. 
 The lowest order is obtained as
\begin{align}
\label{eq:Collect_2}
 \omega^2 = \frac{2 \mu e^2 }{\epsilon_0}
  C(\theta_{\bm q}) |{\bm q}| \; , 
\end{align}
where 
the coefficient $ C(\theta_{\bm q})$ is given by 
\begin{align}
\label{eq:Collect_3}
  C(\theta_{\bm q}) & = \frac{1}{\pi} \int_0^{2\pi} \hspace{-1em}{\rm d}\varphi 
  \frac{1 - \cos^2(\varphi-\theta_{\bm q})}{ 1 + \alpha \cos\varphi}
  \nonumber \\ 
 &
 = \frac{2}{ 1 + \sqrt{1 - \alpha^2}}
 \left[
 1 + \frac{\alpha^2 \sin^2 \theta_{\bm q}}{ \sqrt{1 -\alpha^2}
 (1 + \sqrt{1 - \alpha^2})}
\right].
\end{align}
In this case, the plasma frequency takes
  a minimum, $2/(1+\sqrt{1-\alpha^2})$,
at $\theta_{\bm q}$ = 0 and $\pi$,  and a maximum
, $2/(\sqrt{1-\alpha^2}(1+\sqrt{1-\alpha^2}))$, 
 at $\theta_{\bm q} = \pi/2$.
The plasma frequency in Fig. \ref{Plasmon} reduces to 
  the result in eq.~(\ref{eq:Collect_3}) in the limit of a small $q$. 
 The global behavior including a large $q$ is reported in a separate paper.

\subsection{Screening of Coulomb interaction}
\begin{figure}[tb]
\begin{center}
\includegraphics{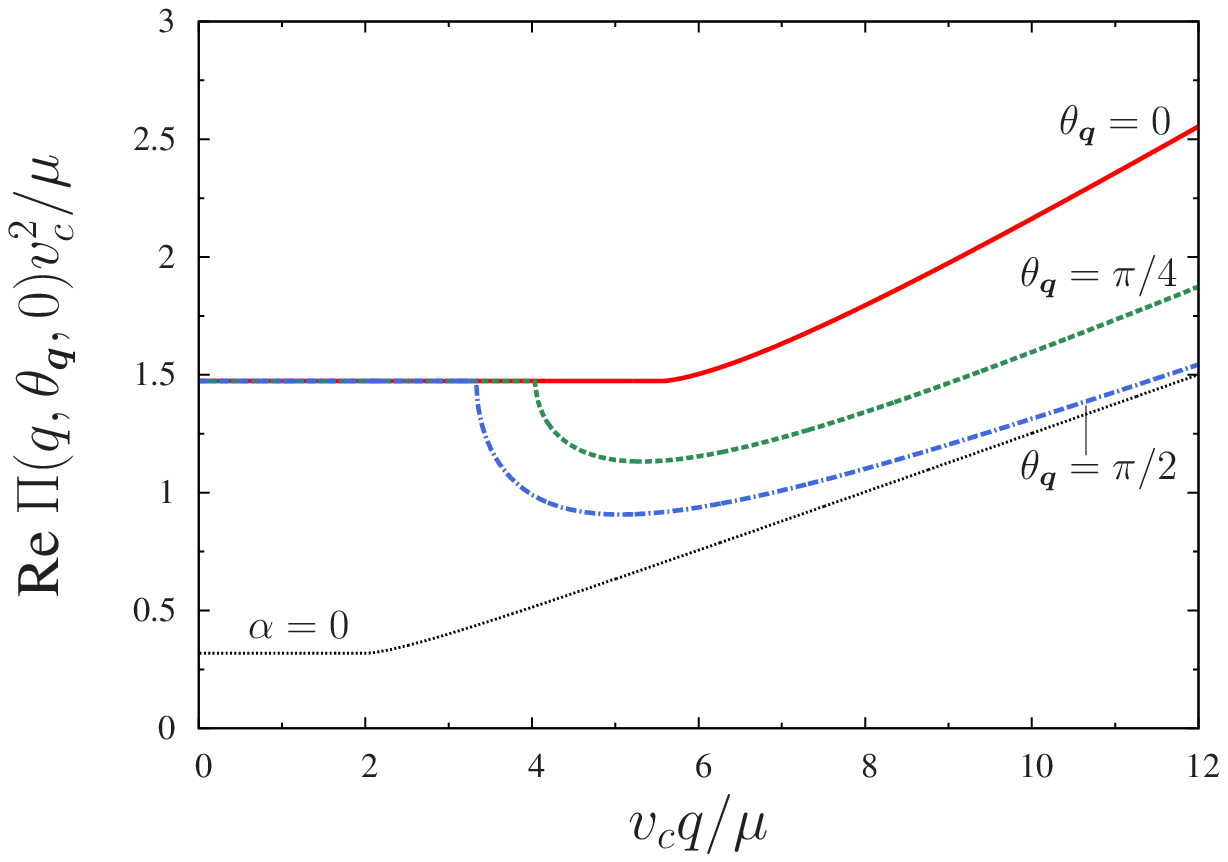}
\end{center}
\caption{
(Color online) 
Real part at $\omega = 0$ as a function of $v_cq/\mu$ with fixed
$\theta _{\bm{q}}= 0, \pi/4, \pi/2$, where the lowest line denotes the result of the isotropic case of graphene.  
}
\label{screening}
\end{figure}

We examine the static screening in the presence of 
Coulomb interaction. In the static case, i.e., $\omega=0$, 
the effective interaction  within the RPA is given by 
\begin{align}
&v^{\rm RPA}(q,\theta_{\bm q},0)
=\frac{v_{\bm q}}{1+v_{\bm q}\Pi(q,\theta_{\bm q},0)}. 
\label{eq:v_RPA}
\end{align}
The real part of $\Pi(q,\theta_{\bm q},0)$ is shown in Fig.~\ref{screening}
with the fixed $\theta_{\bm{q}}=  0, \pi/4$ and $\pi/2$ 
where the isotropic case  is also shown by the dotted line.
There is a cusp at $2 k_{\rm F}(\theta_{\bm{q}})$ (except for 
$\theta_{\bm{q}} = 0$), which is given by 
\begin{align}
 k_{\rm F}(\theta_{\bm{q}}) = \frac{\mu/v_c}
{\sqrt{(1-\alpha^2)(1 - \alpha^2 \cos^2 \theta_{\bm{q}})}}. 
\end{align}
 For $q < 2 k_{\rm F}(\theta_{\bm{q}})$,  $\Pi(q,\theta_{\bm{q}},0)$ remains
 constant as found in the conventional two-dimensional electron gas.
In this regime, the interaction is screened
 as 
\begin{align}
v^{\rm RPA}(q,\theta_{\bm q},0)
\rightarrow 2\pi e^2/\{\epsilon_0(q + q_{\rm TF})\} , 
\end{align}
 where  $q_{\rm TF}$ is the Thomas-Fermi screening constant given by 
$ 2e^2/\{\epsilon_0 (1 - \alpha^2)^{3/2}\}$. 
  For a large $q$, we obtain 
\begin{align}
\Pi(q,\theta_{\bm q},0)
 \rightarrow \frac{q}{8\sqrt{1-\alpha^2 \cos^2 \theta_{\bm{q}}}} ,
\end{align}
which corresponds to  $\mu=0$ in eq.~(\ref{eq:ReP}).  
In the case of a large $q$, the property of the Coulomb interaction 
  remains with the  the dielectric constant replaced  by 
\begin{align}
\epsilon_0+\frac{\pi e^2}{4v_c\sqrt{1-\alpha^2\cos^2\theta_{\bm q}}} .
\end{align}
The dielectric constant takes a maximum (minimum ) 
 at $\theta_{\bm q} = 0  (\pi/2)$
 suggesting a fact that the anisotropy of the velocity 
 gives rise to the enhancement of the  charge response.
 Thus, we found that the effect of the tilted cone also emerges 
  in  the  $\theta_{\bm q}$ dependence of the dielectric constant. 

\section{Conclusions}
We have examined the property of the polarization function 
 of the  massless Dirac particle with  the tilted cone 
 and a finite doping.  Using the tilted Weyl equation,  the dynamical polarization function with the external momentum $\bm{q}$ and frequency $\omega$ is treated analytically,  
  and is applied 
 to calculate the optical conductivity,  plasma frequency, and  screening of the Coulomb interaction.  The tilting of the Dirac cone gives 
 the  following effects on the polarization function,  
which is determined 
 by the intraband and interband excitations.

Several cusps as a function of the frequency, $\omega$,  exist 
 in both the imaginary part and  real part 
of the polarization function  at the characteristic frequencies,  $\omega = \omega_A$, $\omega_B$, and $\omega_+$, depending on the momentum $\bm{q}$ and 
 the tilting parameter $\alpha$. 
Nonmonotonic structures and cusps originate from 
 the tilting and  
 are in contrast to those in the isotropic case ($\alpha = 0$). 
The cusps are understood in terms of the saddle point of the particle-hole excitation energy on the plane of the momentum. 
The tilting effect also appears in the anisotropy of the resonance frequency $\omega_{\rm res}$, which separates the intraband excitation from that of interband one.
 The intensity of ${\rm Im} \Pi(\theta_{\bf q},\omega)$ exhibits a strong asymmetry between the intraband region and the interband region. 
  In the case of  $v_c q/\mu < 1$, ${\rm Im} \Pi(\theta_{\bf q},\omega)$ of the intraband excitation ($\omega < \omega_{\rm res}$) is much larger than that of the interband excitation ($\omega > \omega_{\rm res}$), while the behavior is opposite for $v_c q/\mu >> 1$. 

The optical conductivity, which  is finite above the critical frequency 
 increases continuously from zero  and reaches  a universal value. 
  The plasma frequency for a small $q$, which is proportional to $q^{1/2}$, 
  depends on $\theta_{\bm{q}}$, and  is  symmetric with respect to 
 $\theta_{\bm{q}} = \pi/2$, where the coefficient takes a maximum at 
$\theta_{\bm{q}} = \pi/2$.  
Such an anisotropy  also exists in the screening of the Coulomb interaction, 
where the renormalized dielectric constant becomes largest at $\theta_{\bm{q}} = 0$. 
 The effect of interaction on two  tilted  cones  is reported 
 in a separate paper within RPA.


 We comment on the characteristic energy relevant to  $\alpha$-(BEDT-TTF)$_2$I$_3$. 
It is expected that several cusps obtained in the present paper occur for the energy around the Fermi energy being $\sim$ 5 meV.
However, the observation of these cusps is not straightforward since the neutron scattering is inapplicable owing to the absence of magnetic ordering and the energy scale for the X-ray experiment is too high.
One possible way is to use the electromagnetic wave between microwaves and infrared waves. 
The optical conductivity (Fig.~\ref{ReSgm_omg-dep}), which is obtained in the zero limit of momentum,  also exhibits cusps which are different from those of $\omega_A$, $\omega_B$, and $\omega_+$.  
For the frequency dependence of the optical conductivity, 
the Drude like behavior of the intraband optical conductivity 
is well separated from the interband one since 
the energy for the impurity scattering ($\simeq$ 0.1 meV) is much smaller than the Fermi energy ($\simeq$  5 meV). This is in contrast to the case of the graphene where 
the Drude like contribution  becomes large in addition to the interband one
\cite{Ando_optical}  owing to the large scattering  energy of  10 meV.
\cite{Giesbers2007}
Thus, we may find the behavior showing that the optical conductivity increases continuously from the lower edge of  $  \hbar \omega = 2\mu/(1 +\alpha)$ ( $\sim$ 5 meV) and that the steep increase of the conductivity at the edge depends on $\theta_{\bm{q}} (\not= 0)$. In addition to such a continuous variation, we may expect 
 the effect of finite temperature. Although it is similar to  Fig.~\ref{ReSgm_omg-dep},  it can be distinguished from it since it approaches tangentially to the limiting value. 
Furthermore,  there are also  contributions from  other two bands
   located around  $\sim$ 500 meV below the Fermi energy 
,\cite{Katayama2006JPSJ} where a similar variation with respect
 to the frequency is expected.

Finally, we briefly  mention the case of the zero gap state under pressure 
 in which  the Fermi energy is located on the contact point of the cones (i.e., $\mu \rightarrow 0$). 
Although the effect of tilting still existsts, through the anisotropy of 
$\omega_{\rm res} = (1+ \alpha \cos \theta_{\bm{q}})$, 
   $\Pi({\bm{q},\omega})$  in Fig.~\ref{ImP_region_2pi4} 
 remains finite only in a single region of the interband,
  i.e., $\omega_B$ moves to $\omega_{\rm res}$. 
The optical conductivity becomes constant for both the arbitrary $\omega$  and $\theta_{\bm{q}}$, while the plasma mode vanishes owing to the absence of the Fermi surface, as seen from eq.~(\ref{eq:Collect_2}).

\section*{Acknowledgment}
The authors are thankful to R. Roldan, J.-N. Fuchs, M. O. Goerbig, F. Pi$\acute{\rm e}$chon, and G. Montambaux for fruitful discussions. Y.S. is indebted to the Daiko foundation for the financial aid to the present work. This work was financially supported in part by a Grant-in-Aid for Special Coordination Funds for Promoting Science and Technology (SCF), Scientific Research on Innovative Areas 20110002, and Scientific Research 19740205 from the Ministry of Education, Culture, Sports, Science, and Technology in Japan.

\appendix
\section{Calculation of the Imaginary Part}\label{sec:cal_Im}
By noting that $f(\xi_{s,{\bm k}})\rightarrow \Theta(\mu-\xi_{s,{\bm k}})$ for 
$T\rightarrow 0$ and using 
$(x+i\eta)^{-1}={\cal P}\frac{1}{x}-i\pi\delta(x)$, 
 the imaginary part of eq. (\ref{eq:Polization}) is rewritten as 
\begin{align}
&{\rm Im}\,\Pi({\bm q},\omega) 
 =  \sum_{ss'}\Pi_{ss'}
\nonumber\\
=&\frac{1}{4\pi}\!\int\!\!d{\bm k}
\big|{\bm F}^{\dagger}_{+}({\bm k}){\bm F}_{+}({\bm k}+{\bm q})\big|^2
\Bigl[\Theta(\mu-\xi_{+,{\bm k}})-\Theta(\mu-\xi_{+,{\bm k}+{\bm q}})\Bigr]
\delta\bigl(\omega+\xi_{+,{\bm k}}-\xi_{+,{\bm k}+{\bm q}}\bigr)
& 
\nonumber\\
+&\frac{1}{4\pi}\!\int\!\!d{\bm k}
\big|{\bm F}^{\dagger}_{-}({\bm k})
{\bm F}_{-}({\bm k}+{\bm q})\big|^2
\Bigl[\Theta(\mu-\xi_{-,{\bm k}})-\Theta(\mu-\xi_{-,{\bm k}+{\bm q}})\Bigr]
\delta\bigl(\omega+\xi_{-,{\bm k}}-\xi_{-,{\bm k}+{\bm q}}\bigr)
& 
\nonumber\\
+&\frac{1}{4\pi}\!\int\!\!d{\bm k}
\big|{\bm F}^{\dagger}_{+}({\bm k})
{\bm F}_{-}({\bm k}+{\bm q})\big|^2
\Bigl[\Theta(\mu-\xi_{+,{\bm k}})-\Theta(\mu-\xi_{-,{\bm k}+{\bm q}})\Bigr]
\delta\bigl(\omega+\xi_{+,{\bm k}}-\xi_{-,{\bm k}+{\bm q}}\bigr)
& 
\nonumber\\
+&\frac{1}{4\pi}\!\int\!\!d{\bm k}
\big|{\bm F}^{\dagger}_{-}({\bm k})
{\bm F}_{+}({\bm k}+{\bm q})\big|^2
\Bigl[\Theta(\mu-\xi_{-,{\bm k}})-\Theta(\mu-\xi_{+,{\bm k}+{\bm q}})\Bigr]
\delta\bigl(\omega+\xi_{-,{\bm k}}-\xi_{+,{\bm k}+{\bm q}}\bigr) ,
& 
\label{eq:Im+-}
\end{align}
where 
 ${\rm Im}\,\Pi_{--} = 0$ and 
${\rm Im}\,\Pi_{+-}= 0$ owing to 
$ \mu > 0$. 
${\bm k}+{\bm q} \rightarrow ={\bm k}'$ in  
${\rm Im}\,\Pi_{++}$ and 
${\rm Im}\,\Pi_{-+}$. 
By making use of   eqs. (\ref{eq:disp}) and (\ref{eq:FF2}),
 eq. (\ref{eq:Im+-}) is written as 
\begin{subequations}
\begin{align}
{\rm Im}\,\Pi_{++}
=&\frac{1}{2\pi v_c}\!\int\!\!d{\bm k}
\frac{1}{2}\Bigl[1+\cos(\theta_{\bm k}-\theta_{{\bm k}+{\bm q}})\Bigr]
\Theta\biggl(\frac{\mu}{v_c}-k-\alpha k_x\biggr)
\delta\biggl(\frac{\omega-v_0 q_x}{v_c}+k-|{\bm k}+{\bm q}|\biggr)
 \nonumber \\
-&\frac{1}{2\pi v_c}\!\int\!\!d{\bm k}
\frac{1}{2}\Bigl[1+\cos(\theta_{\bm k}-\theta_{{\bm k}-{\bm q}})\Bigr]
\Theta\biggl(\frac{\mu}{v_c}-k-\alpha k_x\biggr)
\delta\biggl(\frac{\omega-v_0 q_x}{v_c}-k+|{\bm k}-{\bm q}|\biggr) ,
\\
{\rm Im}\,\Pi_{-+}
=&\frac{1}{2\pi v_c}\!\int\!\!d{\bm k}
\frac{1}{2}\Bigl[1-\cos(\theta_{\bm k}-\theta_{{\bm k}-{\bm q}})\Bigr]
\Theta\biggl(k+\alpha k_x-\frac{\mu}{v_c}\biggr)
\delta\biggl(\frac{\omega-v_0 q_x}{v_c}-k-|{\bm k}-{\bm q}|\biggr) .
\end{align}
\label{eq:eqa3}
\end{subequations}
Using the relation 
\begin{align}
\cos(\theta_{\bm k}-\theta_{{\bm k}\pm{\bm q}})
=\frac{{\bm k}\!\cdot\!({\bm k}\pm{\bm q})}{k|{\bm k}\pm{\bm q}|}
=\frac{k\pm q\cos(\theta_{\bm k}-\theta_{\bm q})}{\sqrt{k^2+q^2\pm2kq\cos(\theta_{\bm k}-\theta_{\bm q})}} ,
\end{align}
 with  
$\theta_{\bm k}$ ($\theta_{\bm q}$) being  an angle between 
${\bm k}$ (${\bm q}$) and the $x$-axis, 
eq.~(\ref{eq:eqa3}) is rewritten as  
\begin{subequations}
\begin{align}
{\rm Im}\,\Pi_{++}
=&\frac{1}{2\pi v_c}\!\int_{0}^{\infty}\!\!\!\!\!\!\!kdk
\!\!\int_{-1}^{1}\!\!\frac{du}{\sqrt{1-u^2}}
\left[1+\frac{k+qu}{\sqrt{k^2+q^2+2kqu}}\right]
\nonumber\\
&\times\Theta\biggl(\frac{\mu}{v_c}-k
-\alpha k\left[u\cos\theta_{\bm q}-\sqrt{1-u^2}\sin\theta_{\bm q}\right]\biggr)
\delta\Bigl(g_1(u)\Bigr) 
\label{eq:g1}
 \nonumber \\
-&\frac{1}{2\pi v_c}\!\int_{0}^{\infty}\!\!\!\!\!\!\!kdk
\!\!\int_{-1}^{1}\!\!\frac{du}{\sqrt{1-u^2}}
\left[1+\frac{k-qu}{\sqrt{k^2+q^2-2kqu}}\right] 
 \nonumber \\ 
&\times\Theta\biggl(\frac{\mu}{v_c}-k
-\alpha k\left[u\cos\theta_{\bm q}-\sqrt{1-u^2}\sin\theta_{\bm q}\right]\biggr)
\delta\Bigl(g_2(u)\Bigr) ,
\\
{\rm Im}\,\Pi_{-+}
=&\frac{1}{2\pi v_c}\!\int_{0}^{\infty}\!\!\!\!\!\!\!kdk
\!\!\int_{-1}^{1}\!\!\frac{du}{\sqrt{1-u^2}}
\left[1-\frac{k-qu}{\sqrt{k^2+q^2-2kqu}}\right]
\nonumber\\
&\times\Theta\biggl(k
+\alpha k\left[u\cos\theta_{\bm q}-\sqrt{1-u^2}\sin\theta_{\bm q}\right]-\frac{\mu}{v_c}\biggr)
\delta\Bigl(g_3(u)\Bigr) ,
\label{eq:g3}
\end{align}
\end{subequations}
 where 
$\varphi=\theta_{\bm k}-\theta_{\bm q}$, 
$\nu=\frac{\omega-v_0 q_x}{v_c}=\frac{\omega}{v_c}-\alpha q\cos\theta_{\bm q}$,  where $g_1,g_2$, and $g_3$ are given by  
\begin{subequations}
\begin{align}
&g_1(u)=\nu+k-\sqrt{k^2+q^2+2kqu} \; , 
\\
&g_2(u)=\nu-k+\sqrt{k^2+q^2-2kqu} \; ,
\\
&g_3(u)=\nu-k-\sqrt{k^2+q^2-2kqu} \; .
\end{align}
\end{subequations} 
By noting  the necessary condition for $g_i(u)=0$  ( $-1\leq u\leq 1$), 
   the $\delta$ function   is evaluated as  
\begin{align}
g_1(u)=0
&\rightarrow 
\Theta(q-\nu)\Theta\left(k-\frac{q-\nu}{2}\right) ,
\\
g_2(u)=0
&\rightarrow 
\Theta(q-\nu)\Theta\left(k-\frac{q+\nu}{2}\right) ,
\\
g_3(u)=0
&\rightarrow 
\Theta(\nu-q)\Theta\left(k-\frac{\nu-q}{2}\right)\Theta\left(\frac{\nu+q}{2}-k\right) .
\end{align}
 The integration  with respect to $u$  is performed as
\begin{subequations}
\begin{align}
{\rm Im}\,\Pi_{++}
=&\frac{1}{2\pi v_c}\!\int_{0}^{\infty}\!\!\!\!\!\!\!dk
\sqrt{\frac{(2k+\nu)^2-q^2}{q^2-\nu^2}}
\Theta(q-\nu)\Theta\left(k-\frac{q-\nu}{2}\right)
\nonumber\\
&\times\Theta\Biggl(\frac{2\mu+\omega}{v_c}-\left(q+\alpha\nu\cos\theta_{\bm q}\right)
\frac{2k+\nu}{q}
+\alpha\sin\theta_{\bm q}\sqrt{q^2-\nu^2}
\sqrt{\left(\frac{2k+\nu}{q}\right)^{\!\!\!2}-1}\Biggr)
\label{eq:k1}
\nonumber \\
-&\frac{1}{2\pi v_c}\!\int_{0}^{\infty}\!\!\!\!\!\!\!dk
\sqrt{\frac{(2k-\nu)^2-q^2}{q^2-\nu^2}}
\Theta(q-\nu)\Theta\left(k-\frac{q+\nu}{2}\right)
\nonumber\\
&\times\Theta\Biggl(\frac{2\mu-\omega}{v_c}-\left(q+\alpha\nu\cos\theta_{\bm q}\right)
\frac{2k-\nu}{q}
+\alpha\sin\theta_{\bm q}\sqrt{q^2-\nu^2}
\sqrt{\left(\frac{2k-\nu}{q}\right)^{\!\!\!2}-1}\Biggr)  \; ,
\\
{\rm Im}\,\Pi_{-+}
=&\frac{1}{2\pi v_c}\!\int_{0}^{\infty}\!\!\!\!\!\!\!dk
\sqrt{\frac{q^2-(2k-\nu)^2}{\nu^2-q^2}}
\Theta(\nu-q)\Theta\left(k-\frac{\nu-q}{2}\right)\Theta\left(\frac{\nu+q}{2}-k\right)
\nonumber\\
&\times\Theta\Biggl(-\frac{2\mu-\omega}{v_c}+\left(q+\alpha\nu\cos\theta_{\bm q}\right)
\frac{2k-\nu}{q}
-\alpha\sin\theta_{\bm q}\sqrt{\nu^2-q^2}
\sqrt{1-\left(\frac{2k-\nu}{q}\right)^{\!\!\!2}}\Biggr) \; .
\label{eq:k3}
\end{align}
\end{subequations}
Replacing $k$ as $k'=\frac{2k+\nu}{q}$  and  
 $k'=\frac{2k-\nu}{q}$ in eqs. (\ref{eq:k1}) and (\ref{eq:k3}), one obtains 
\begin{subequations}
\begin{align}
{\rm Im}\,\Pi_{++}
=&\frac{1}{4\pi v_c}\frac{q^2\Theta(q-\nu)}{\sqrt{q^2-\nu^2}}
\!\int_{\frac{\nu}{q}}^{\infty}\!\!\!\!\!\!\!dk'
\sqrt{k^{'2}-1}\Theta\left(k'-1\right)
\nonumber\\
&\times\Theta\biggl(\frac{2\mu+\omega}{v_c}-\left(q+\alpha\nu\cos\theta_{\bm q}\right)k'
+\alpha\sin\theta_{\bm q}\sqrt{q^2-\nu^2}\sqrt{k^{'2}-1}\biggr)
\label{eq:k1'}
\nonumber \\
-&\frac{1}{4\pi v_c}\frac{q^2\Theta(q-\nu)}{\sqrt{q^2-\nu^2}}
\!\int_{-\frac{\nu}{q}}^{\infty}\!\!\!\!\!\!\!dk'
\sqrt{k^{'2}-1}\Theta\left(k'-1\right)
\nonumber\\
&\times\Theta\biggl(\frac{2\mu-\omega}{v_c}-\left(q+\alpha\nu\cos\theta_{\bm q}\right)k'
+\alpha\sin\theta_{\bm q}\sqrt{q^2-\nu^2}\sqrt{k^{'2}-1}\biggr) \; ,
\\
{\rm Im}\,\Pi_{-+}
=&\frac{1}{4\pi v_c}\frac{q^2\Theta(\nu-q)}{\sqrt{\nu^2-q^2}}
\!\int_{-\frac{\nu}{q}}^{\infty}\!\!\!\!\!\!\!dk'
\sqrt{1-k^{'2}}\Theta\left(k'+1\right)\Theta\left(1-k'\right)
\nonumber\\
&\times\Theta\biggl(-\frac{2\mu-\omega}{v_c}+\left(q+\alpha\nu\cos\theta_{\bm q}\right)k'
-\alpha\sin\theta_{\bm q}\sqrt{\nu^2-q^2}\sqrt{1-k^{'2}}\biggr) \; .
\label{eq:k3'}
\end{align}
\end{subequations}
The step function in the above equation for $(q>0,\omega >0)$ 
 gives the following condition. 

When  $q>\nu$, 
 $q+\alpha\nu\cos\theta_{\bm q}>0$ for 
 $0<\theta_{\bm q}<\pi$. 
 When $\nu>q$, 
$q+\alpha\nu\cos\theta_{\bm q}>0$
 for $0<\theta_{\bm q}<\frac{\pi}{2}$ , and 
 $q+\alpha\nu\cos\theta_{\bm q}<0$
 with  
$\omega>-\frac{v_c(1-\alpha^2\cos^2\theta_{\bm q})}{\alpha\cos\theta_{\bm q}}$
 for $\frac{\pi}{2}<\theta_{\bm q}<\pi$. 
Thus, eqs.~(\ref{eq:k1'}) 
and (\ref{eq:k3'}) are rewritten as
\begin{subequations}
\begin{align}
{\rm Im}\,\Pi_{++}
=&\frac{1}{4\pi v_c}\frac{q^2\Theta(q-\nu)}{\sqrt{q^2-\nu^2}}
\Theta(x_1-1)
\!\int_{1}^{x_1^{+}}\!\!\!\!\!\!\!\!dk'
\sqrt{k^{'2}-1}
\nonumber\\
+&\frac{1}{4\pi v_c}\frac{q^2\Theta(q-\nu)}{\sqrt{q^2-\nu^2}}
\Theta\left(\frac{2\mu+\omega}{v_c}-\sqrt{U}\right)
\Theta(1-x_1)
\!\int_{x_1^{-}}^{x_1^{+}}\!\!\!\!\!\!\!\!dk'
\sqrt{k^{'2}-1}
\label{eq:kk11}
\nonumber \\
-&\frac{1}{4\pi v_c}\frac{q^2\Theta(q-\nu)}{\sqrt{q^2-\nu^2}}
\Theta(x_2-1)
\!\int_{1}^{x_2^{+}}\!\!\!\!\!\!\!\!dk'
\sqrt{k^{'2}-1}
\nonumber\\
-&\frac{1}{4\pi v_c}\frac{q^2\Theta(q-\nu)}{\sqrt{q^2-\nu^2}}
\Theta\left(\frac{2\mu-\omega}{v_c}-\sqrt{U}\right)
\Theta(1-x_2)
\!\int_{x_2^{-}}^{x_2^{+}}\!\!\!\!\!\!\!\!dk'
\sqrt{k^{'2}-1} \; ,
\\
{\rm Im}\,\Pi_{-+}
=&\frac{1}{4\pi v_c}\frac{q^2\Theta(\nu-q)}{\sqrt{\nu^2-q^2}}
\Theta(1-|x_2|)\Theta(q+\alpha\nu\cos\theta_{\bm q})
\!\int_{x_2^{+}}^{1}\!\!\!\!\!dk'
\sqrt{1-k^{'2}}
\nonumber\\
+&\frac{1}{4\pi v_c}\frac{q^2\Theta(\nu-q)}{\sqrt{\nu^2-q^2}}
\Theta(1-|x_2|)\Theta(-q-\alpha\nu\cos\theta_{\bm q})
\!\int_{-1}^{x_2^{-}}\!\!\!\!\!\!\!\!dk'
\sqrt{1-k^{'2}}
\nonumber\\
+&\frac{1}{4\pi v_c}\frac{q^2\Theta(\nu-q)}{\sqrt{\nu^2-q^2}}
\Theta\left(\sqrt{U}-\frac{\omega-2\mu}{v_c}\right)
\Theta(-1-x_2)\Theta(q+\alpha\nu\cos\theta_{\bm q})
\!\int_{x_2^{+}}^{1}\!\!\!\!\!dk'
\sqrt{1-k^{'2}}
\nonumber\\
+&\frac{1}{4\pi v_c}\frac{q^2\Theta(\nu-q)}{\sqrt{\nu^2-q^2}}
\Theta\left(\sqrt{U}-\frac{\omega-2\mu}{v_c}\right)
\Theta(x_2-1)\Theta(-q-\alpha\nu\cos\theta_{\bm q})
\!\int_{x_2^{+}}^{1}\!\!\!\!\!dk'
\sqrt{1-k^{'2}}
\nonumber\\
+&\frac{1}{4\pi v_c}\frac{q^2\Theta(\nu-q)}{\sqrt{\nu^2-q^2}}
\Theta\left(\sqrt{U}-\frac{\omega-2\mu}{v_c}\right)
\Theta(-1-x_2)\Theta(q+\alpha\nu\cos\theta_{\bm q})
\!\int_{-1}^{x_2^{-}}\!\!\!\!\!\!\!\!dk'
\sqrt{1-k^{'2}}
\nonumber\\
+&\frac{1}{4\pi v_c}\frac{q^2\Theta(\nu-q)}{\sqrt{\nu^2-q^2}}
\Theta\left(\sqrt{U}-\frac{\omega-2\mu}{v_c}\right)
\Theta(x_2-1)\Theta(-q-\alpha\nu\cos\theta_{\bm q})
\!\int_{-1}^{x_2^{-}}\!\!\!\!\!\!\!\!dk'
\sqrt{1-k^{'2}}
\nonumber\\
+&\frac{1}{4\pi v_c}\frac{q^2\Theta(\nu-q)}{\sqrt{\nu^2-q^2}}
\Theta\left(\frac{\omega-2\mu}{v_c}-\sqrt{U}\right)
\!\int_{-1}^{1}\!\!\!\!\!dk'
\sqrt{1-k^{'2}} \; .
\label{eq:kk33}
\end{align}
\end{subequations}


\appendix
\section{Expression of the Imaginary Part} \label{Exp_Im}

Performing the $k'$-integration in eqs.~(\ref{eq:kk11}) 
 and (\ref{eq:kk33}), 
eqs.~(\ref{eq:Pi_++}) and (\ref{eq:Pi_-+}) are calculated as follows:

\begin{subequations}
\begin{align}
&
\Pi''_{\rm 1A}=
f(q,\nu)
\Big[
G_{>}(x_1^{+})+{\rm sgn}(x_1-1)G_{>}(x_1^{-})-G_{>}(x_2^{+})-{\rm sgn}(x_2-1)G_{>}(x_2^{-})
\Bigr]
\Theta(\omega_{\rm res}-\omega)
\Theta(\omega_{\rm A}-\omega) , 
\label{eq:Pi_17a}
\\
&
\Pi''_{\rm 2A}=
f(q,\nu)
\Big[
G_{>}(x_1^{+})+{\rm sgn}(x_1-1)G_{>}(x_1^{-})
\Bigr]
\Theta(\omega_{\rm res}-\omega)
\Theta(\omega-\omega_{\rm A})
\Theta(\omega-\omega_{+}) ,
\label{eq:Pi_17b}
\\
&
\Pi''_{\rm 3A}=
0\cdot
\Theta(\omega_{+}-\omega)
\label{eq:ImP_A} ,
\end{align}
\end{subequations}

\begin{subequations}
\begin{align}
&
\Pi''_{\rm 1B}=
0\cdot
\Theta(\omega-\omega_{\rm res})
\Theta(\omega_{\rm A}-\omega) ,
\label{eq:Pi_19a}
\\
&
\Pi''_{\rm 2B}=
f(q,\nu)
\Big[
\pi+{\rm sgn}(x_2-1)G_{<}(x_2^{+})+{\rm sgn}(x_2+1)G_{<}(-x_2^{-})
\Bigr]
\Theta(\omega-\omega_{\rm res})
\Theta(\omega_{\rm B}-\omega)
\Theta(\omega-\omega_{\rm A}) ,
\label{eq:Pi_19b}
\\
&
\Pi''_{\rm 3B}=
f(q,\nu)
2\pi
\Theta(\omega-\omega_{\rm B}) .
\label{eq:ImP_B}
\end{align}
\end{subequations}

In the above equations, we define 
\begin{align}
&
f(q,\nu)=\frac{1}{8\pi v_c}\frac{q^2}{\sqrt{|q^2-\nu^2|}} ,
\label{eq:f}
\\
&
G_{>}(x)=x\sqrt{x^{2}-1}-{\rm arccosh}(x)\quad{\rm for}\quad x>1 ,
\label{eq:Ga}
\\
&
G_{<}(x)=x\sqrt{1-x^{2}}-\arccos(x)\quad{\rm for}\quad |x|<1 , 
\label{eq:Gb}
\end{align}
where
\begin{align}
&
\nu=\frac{\omega}{v_c}-\alpha q \cos\theta_{\bm q} ,
\label{eq:nu}
\\
&
U=(1-\alpha^2)(q^2-\nu^2)+\left(\frac{\omega}{v_c}\right)^{\!\!2} ,
\label{eq:U}
\\
&
x_1=\frac{2\mu+\omega}{v_c(q+\alpha\nu\cos\theta_{\bm q})} ,
\label{eq:x1}
\\
&
x_2=\frac{2\mu-\omega}{v_c(q+\alpha\nu\cos\theta_{\bm q})} ,
\label{eq:x2}
\\
&
x_1^{\pm}=\frac{2\mu+\omega}{v_c U}(q+\alpha\nu\cos\theta_{\bm q})
\pm\frac{\alpha|\sin\theta_{\bm q}|}{U}\sqrt{\bigl(q^2-\nu^2\bigr)
\left\{\left(\frac{2\mu+\omega}{v_c}\right)^{\!\!2}-U\right\}} ,
\label{eq:x1pm}
\\
&
x_2^{\pm}=\frac{2\mu-\omega}{v_c U}|q+\alpha\nu\cos\theta_{\bm q}|
\pm\frac{\alpha|\sin\theta_{\bm q}|}{U}\sqrt{\bigl(q^2-\nu^2\bigr)
\left\{\left(\frac{2\mu-\omega}{v_c}\right)^{\!\!2}-U\right\}} ,
\label{eq:x2pm}
\\
&
\frac{
\omega_{+}(\theta_{\bm q})
}{\mu}
=\alpha\frac{v_c q}{\mu}\cos\theta_{\bm q}-\frac{2}{1-\alpha^2}
+\sqrt{\left(\frac{v_c q}{\mu}\right)^{\!\!2}-\frac{4\alpha\frac{v_c q}{\mu}\cos\theta_{\bm q}}{1-\alpha^2}
+\left(\frac{2\alpha}{1-\alpha^2}\right)^{\!\!2}} , 
\label{eq:omega_+}
\\
&
\frac{
\omega_{\rm A(B)}(\theta_{\bm q})
}{\mu}
=\alpha\frac{v_c q}{\mu}\cos\theta_{\bm q}+\frac{2}{1-\alpha^2}
-(+)\sqrt{\left(\frac{v_c q}{\mu}\right)^{\!\!2}+\frac{4\alpha\frac{v_c q}{\mu}\cos\theta_{\bm q}}{1-\alpha^2}
+\left(\frac{2\alpha}{1-\alpha^2}\right)^{\!\!2}} . 
\label{eq:omega_AB}
\end{align}

\appendix
\section{Expression of the Real Part}\label{sec:ex_Real}

The  real part is calculated using eqs.~(\ref{eq:KK}) and (\ref{eq:sym})
 in which the  semianalytical calculation is performed
 by dividing the part  into four regions, i.e.,
$({\rm I})\,0<v_cq/\mu < 1/(1 + \alpha |\cos \theta_{\bm{q}}|)$, 
$({\rm I\hspace{-0.1em}I})\,1/(1 + \alpha |\cos \theta_{\bm{q}}|)
<v_c q/\mu<1/(1-\alpha|\cos\theta_{\bm{q}}|)$, 
$({\rm I\hspace{-0.1em}I\hspace{-0.1em}I})\,1/(1-\alpha|\cos\theta_{\bm{q}}|)
<v_c q/\mu<2/(\sqrt{1-\alpha^2}\sqrt{1-\alpha^2\cos^2\theta_{\bm{q}}})$, 
and 
$({\rm I\hspace{-0.1em}V})\,2/(\sqrt{1-\alpha^2}\sqrt{1-\alpha^2\cos^2\theta_{\bm{q}}})
<v_c q/\mu$.
\begin{align}
{\rm Re\,}\Pi(q,\theta_{\bm q},\omega)
=\Pi'_{\rm 3B}
+\sum_{\zeta={\rm I,I\hspace{-0.1em}I,I\hspace{-0.1em}I\hspace{-0.1em}I,I\hspace{-0.1em}V}}\Pi'_{\zeta} ,
\end{align}
where the first term denotes the the contribution of  $\Pi''_{\rm 3B}$,
while $\Pi'_{\rm I}, ..., \Pi'_{\rm I\hspace{-0.1em}V}$ 
are the semianalytical expressions including the integral. 

The final result is given by
\begin{subequations}
\begin{align}
\label{eq:ReP}
\Pi'_{\rm 3B}=
&
f(q,\nu)\Theta(\nu^2-q^2)
\log\Bigg|
\frac{q\bigl\{\omega-\omega_{\rm B}(\theta_{\bm q})\bigr\}}
{\nu\bigl\{\omega-\omega_{\rm B}(\theta_{\bm q})\bigr\}+q^2-\nu^2
+\sqrt{(\nu^2-q^2)\left\{\bigl(\omega_{\rm B}(\theta_{\bm q})-\alpha q\cos\theta_{\bm q}\bigr)^2-q^2\right\}}}
\nonumber\\
&\times
\frac{q\bigl\{\omega+\omega_{\rm B}(\pi+\theta_{\bm q})\bigr\}}
{\nu\bigl\{\omega+\omega_{\rm B}(\pi+\theta_{\bm q})\bigr\}+q^2-\nu^2
+\sqrt{(\nu^2-q^2)\left\{\bigl(\omega_{\rm B}(\pi+\theta_{\bm q})+\alpha q\cos\theta_{\bm q}\bigr)^2-q^2\right\}}}
\Bigg|
\nonumber\\
&+
f(q,\nu)\Theta(q^2-\nu^2)
\Biggl\{
\arcsin\frac{\nu\bigl\{\omega-\omega_{\rm B}(\theta_{\bm q})\bigr\}+q^2-\nu^2}
{q\big|\omega-\omega_{\rm B}(\theta_{\bm q})\big|}
+\arcsin\frac{\nu\bigl\{\omega+\omega_{\rm B}(\pi+\theta_{\bm q})\bigr\}+q^2-\nu^2}
{q\big|\omega+\omega_{\rm B}(\pi+\theta_{\bm q})\big|}
\Biggr\} ,
\\
\Pi'_{\rm I}=
&
\biggl\{
I^{+}_{\rm 1A}\bigl[\omega_{\rm res}(\theta_{\bm q}),0\bigr]
+I^{+}_{\rm 2B}\bigl[\omega_{\rm A}(\theta_{\bm q}),\omega_{\rm B}(\theta_{\bm q})\bigr]
+\frac{1}{\pi}{\rm Im}\,\Pi(\theta_{\bm q},\omega)
\log
\bigg| \frac{\omega-\omega_{\rm res}(\theta_{\bm q})}{\omega}
\frac{\omega-\omega_{\rm B}(\theta_{\bm q})}{\omega-\omega_{\rm A}(\theta_{\bm q})}\bigg|
\biggr.
\nonumber\\
\biggl.
&
+I^{-}_{\rm 1A}\bigl[\omega_{\rm res}(\pi-\theta_{\bm q}),0\bigr]
+I^{-}_{\rm 2B}\bigl[\omega_{\rm A}(\pi-\theta_{\bm q}),\omega_{\rm B}(\pi-\theta_{\bm q})\bigr]
\biggr\}
\nonumber\\
&\times
\Theta\left(\frac{\mu}{1+\alpha|\cos\theta_{\bm q}|}-v_c q\right) ,
\\
\Pi'_{\rm I\hspace{-0.1em}I}=
&
\Biggl[
\biggl\{
I^{+}_{\rm 1A}\bigl[\omega_{\rm A}(\theta_{\bm q}),0\bigr]
+I^{+}_{\rm 2A}\bigl[\omega_{\rm res}(\theta_{\bm q}),\omega_{\rm A}(\theta_{\bm q})\bigr]
+I^{+}_{\rm 2B}\bigl[\omega_{\rm res}(\theta_{\bm q}),\omega_{\rm B}(\theta_{\bm q})\bigr]
+\frac{1}{\pi}{\rm Im}\,\Pi(\theta_{\bm q},\omega)
\log
\bigg| \frac{\omega-\omega_{\rm B}(\theta_{\bm q})}{\omega} \bigg|
\biggr.
\Biggr.
\nonumber\\
\Biggl.
\biggl.
&
+I^{-}_{\rm 1A}\bigl[\omega_{\rm res}(\pi-\theta_{\bm q}),0\bigr]
+I^{-}_{\rm 2B}\bigl[\omega_{\rm A}(\pi-\theta_{\bm q}),\omega_{\rm B}(\pi-\theta_{\bm q})\bigr]
\biggr\}
\Theta\left(\frac{\pi}{2}-\theta_{\bm q}\right)
\Biggr.
\nonumber\\
&+
\Biggl.
\biggl\{
I^{+}_{\rm 1A}\bigl[\omega_{\rm res}(\theta_{\bm q}),0\bigr]
+I^{+}_{\rm 2B}\bigl[\omega_{\rm A}(\theta_{\bm q}),\omega_{\rm B}(\theta_{\bm q})\bigr]
+\frac{1}{\pi}{\rm Im}\,\Pi(\theta_{\bm q},\omega)
\log
\bigg| \frac{\omega-\omega_{\rm res}(\theta_{\bm q})}{\omega}
\frac{\omega-\omega_{\rm B}(\theta_{\bm q})}{\omega-\omega_{\rm A}(\theta_{\bm q})}\bigg|
\biggr.
\Biggr.
\nonumber\\
\Biggl.
\biggl.
&
+I^{-}_{\rm 1A}\bigl[\omega_{\rm A}(\pi-\theta_{\bm q}),0\bigr]
+I^{-}_{\rm 2A}\bigl[\omega_{\rm res}(\pi-\theta_{\bm q}),\omega_{\rm A}(\pi-\theta_{\bm q})\bigr]
+I^{-}_{\rm 2B}\bigl[\omega_{\rm res}(\pi-\theta_{\bm q}),\omega_{\rm B}(\pi-\theta_{\bm q})\bigr]
\biggr\}
\Theta\left(\theta_{\bm q}-\frac{\pi}{2}\right)
\Biggr]
\nonumber\\
&\times
\Theta\left(\frac{\mu}{1-\alpha|\cos\theta_{\bm q}|}-v_c q\right)
\Theta\left(v_c q-\frac{\mu}{1+\alpha|\cos\theta_{\bm q}|}\right) ,
\\
\Pi'_{\rm I\hspace{-0.1em}I\hspace{-0.1em}I}=
&
\biggl\{
I^{+}_{\rm 1A}\bigl[\omega_{\rm A}(\theta_{\bm q}),0\bigr]
+I^{+}_{\rm 2A}\bigl[\omega_{\rm res}(\theta_{\bm q}),\omega_{\rm A}(\theta_{\bm q})\bigr]
+I^{+}_{\rm 2B}\bigl[\omega_{\rm res}(\theta_{\bm q}),\omega_{\rm B}(\theta_{\bm q})\bigr]
+\frac{1}{\pi}{\rm Im}\,\Pi(\theta_{\bm q},\omega)
\log
\bigg| \frac{\omega-\omega_{\rm B}(\theta_{\bm q})}{\omega} \bigg|
\biggr.
\nonumber\\
\biggl.
&
+I^{-}_{\rm 1A}\bigl[\omega_{\rm A}(\pi-\theta_{\bm q}),0\bigr]
+I^{-}_{\rm 2A}\bigl[\omega_{\rm res}(\pi-\theta_{\bm q}),\omega_{\rm A}(\pi-\theta_{\bm q})\bigr]
+I^{-}_{\rm 2B}\bigl[\omega_{\rm res}(\pi-\theta_{\bm q}),\omega_{\rm B}(\pi-\theta_{\bm q})\bigr]
\biggr\}
\nonumber\\
&\times
\Theta\left(\frac{2\mu}{\sqrt{(1-\alpha^2)(1-\alpha^2\cos^2\theta_{\bm q})}}-v_c q\right)
\Theta\left(v_c q-\frac{\mu}{1-\alpha|\cos\theta_{\bm q}|}\right) ,
\\
\Pi'_{\rm I\hspace{-0.1em}V}=
&
\biggl\{
I^{+}_{\rm 2A}\bigl[\omega_{\rm res}(\theta_{\bm q}),\omega_{+}(\theta_{\bm q})\bigr]
+I^{+}_{\rm 2B}\bigl[\omega_{\rm res}(\theta_{\bm q}),\omega_{\rm B}(\theta_{\bm q})\bigr]
+\frac{1}{\pi}{\rm Im}\,\Pi(\theta_{\bm q},\omega)
\log
\bigg| \frac{\omega-\omega_{\rm B}(\theta_{\bm q})}{\omega-\omega_{+}(\theta_{\bm q})} \bigg|
\biggr.
\nonumber\\
\biggl.
&
+I^{-}_{\rm 2A}\bigl[\omega_{\rm res}(\pi-\theta_{\bm q}),\omega_{+}(\pi-\theta_{\bm q})\bigr]
+I^{-}_{\rm 2B}\bigl[\omega_{\rm res}(\pi-\theta_{\bm q}),\omega_{\rm B}(\pi-\theta_{\bm q})\bigr]
\biggr\}
\nonumber\\
&\times
\Theta\left(v_c q-\frac{2\mu}{\sqrt{(1-\alpha^2)(1-\alpha^2\cos^2\theta_{\bm q})}}\right).
\end{align}
\end{subequations}
Here, we denote the functions defined by the integrals
\begin{subequations}
\begin{align}
&
I^{+}_{\rm 1A/2A}\bigl[a,b\bigr]=
\frac{q}{\pi}\int
_{{\rm arccos}\frac{a-\alpha q\cos\theta_{\bm q}}{q}}
^{{\rm arccos}\frac{b-\alpha q\cos\theta_{\bm q}}{q}}
\hspace{-1em}dy\,\,
\frac{H_{\rm 1A/2A}(q\cos y+\alpha q\cos\theta_{\bm q})-\sin y{\rm Im}\,\Pi(\theta_{\bm q},\omega)}
{q(\cos y+\alpha\cos\theta_{\bm q})-\omega},
\\
&
I^{+}_{\rm 2B}\bigl[a,b\bigr]=
\frac{q}{\pi}\int
_{{\rm arccosh}\frac{a-\alpha q\cos\theta_{\bm q}}{q}}
^{{\rm arccosh}\frac{b-\alpha q\cos\theta_{\bm q}}{q}}
\hspace{-1em}dy\,\,
\frac{H_{\rm 2B}(q\cosh y+\alpha q\cos\theta_{\bm q})-\sinh y{\rm Im}\,\Pi(\theta_{\bm q},\omega)}
{q(\cosh y+\alpha\cos\theta_{\bm q})-\omega},
\\
&
I^{-}_{\rm 1A/2A}\bigl[a,b\bigr]=
\frac{q}{\pi}\int
_{{\rm arccos}\frac{a+\alpha q\cos\theta_{\bm q}}{q}}
^{{\rm arccos}\frac{b+\alpha q\cos\theta_{\bm q}}{q}}
\hspace{-1em}dy\,\,
\frac{H_{\rm 1A/2A}(q\cos y-\alpha q\cos\theta_{\bm q})}
{q(\cos y-\alpha\cos\theta_{\bm q})+\omega},
\\
&
I^{-}_{\rm 2B}\bigl[a,b\bigr]=
\frac{q}{\pi}\int
_{{\rm arccosh}\frac{a+\alpha q\cos\theta_{\bm q}}{q}}
^{{\rm arccosh}\frac{b+\alpha q\cos\theta_{\bm q}}{q}}
\hspace{-1em}dy\,\,
\frac{H_{\rm 2B}(q\cosh y-\alpha q\cos\theta_{\bm q})}
{q(\cosh y-\alpha\cos\theta_{\bm q})+\omega},
\end{align}
\end{subequations}
where
\begin{subequations}
\begin{align}
&
H_{\rm 1A}(\omega)=
\frac{q}{16\pi}
\Bigl[
G_{>}\left(x_{1}^{+}(\omega)\right)
+{\rm sgn}\left(x_{1}(\omega)-1\right)G_{>}\left(x_{1}^{-}(\omega)\right)
-G_{>}\left(x_{2}^{+}(\omega)\right)
-{\rm sgn}\left(x_{2}(\omega)-1\right)G_{>}\left(x_{2}^{-}(\omega)\right)
\Bigr],
\\
&
H_{\rm 2A}(\omega)=
\frac{q}{16\pi}
\Bigl[
G_{>}\left(x_{1}^{+}(\omega)\right)
+{\rm sgn}\left(x_{1}(\omega)-1\right)G_{>}\left(x_{1}^{-}(\omega)\right)
\Bigr],
\\
&
H_{\rm 2B}(\omega)=
\frac{q}{16\pi}
\Bigl[
\pi
+{\rm sgn}\left(x_{2}(\omega)-1\right)G_{<}\left(x_{2}^{+}(\omega)\right)
-{\rm sgn}\left(x_{2}(\omega)+1\right)G_{<}\left(-x_{2}^{-}(\omega)\right)
\Bigr].
\end{align}
\end{subequations}




\end{document}